\documentstyle[11pt]{article}
\def\lvm{\leavevmode\hbox to\parindent{\hfill}}

\def\deg{\mathop{\rm deg}\nolimits}
\def\hat{\widehat}

\def\N#1{N\!=\!#1}
\def\SL#1{s\ell(#1)}

\font\ssfbig=cmss10 scaled\magstephalf
\font\ssfscr=cmss8 scaled\magstephalf
\font\ssfscrscr=cmss8
\newfam\ssffam
\textfont\ssffam=\ssfbig
\scriptfont\ssffam=\ssfscr
\scriptscriptfont\ssffam=\ssfscrscr
\def\ssf{\fam\ssffam}

\makeatletter
\@addtoreset{equation}{section}
\newdimen\normalarrayskip
\newdimen\minarrayskip
\normalarrayskip\baselineskip
\minarrayskip\jot
\newif\ifold \oldtrue \def\new{\oldfalse}
\def\arraymode{\ifold\relax\else\displaystyle\fi}

\def\@arrayskip{\ifold\baselineskip\z@\lineskip\z@
  \else
  \baselineskip\minarrayskip\lineskip2\minarrayskip\fi}
\def\@arrayclassz{\ifcase \@lastchclass \@acolampacol \or
\@ampacol \or \or \or \@addamp \or
 \@acolampacol \or \@firstampfalse \@acol \fi
\edef\@preamble{\@preamble
 \ifcase \@chnum
  \hfil$\relax\arraymode\@sharp$\hfil
  \or $\relax\arraymode\@sharp$\hfil
  \or \hfil$\relax\arraymode\@sharp$\fi}}
\def\@array[#1]#2{\setbox\@arstrutbox=\hbox{\vrule
  height\arraystretch \ht\strutbox
  depth\arraystretch \dp\strutbox
  width\z@}\@mkpream{#2}\edef\@preamble{\halign \noexpand\@halignto
\bgroup \tabskip\z@ \@arstrut \@preamble \tabskip\z@ \cr}%
\let\@startpbox\@@startpbox \let\@endpbox\@@endpbox
 \if #1t\vtop \else \if#1b\vbox \else \vcenter \fi\fi
 \bgroup \let\par\relax
 \let\@sharp##\let\protect\relax
 \@arrayskip\@preamble}
\def\l@section#1#2{\addpenalty{\@secpenalty} \addvspace{.4em plus 1pt}
\@tempdima 1.5em \begingroup
 \parindent \z@ \rightskip \@pnumwidth
 \parfillskip -\@pnumwidth
 \small\sf \leavevmode \advance\leftskip\@tempdima
 \hskip -\leftskip #1\nobreak\hfil
\nobreak\hbox to\@pnumwidth{\hss #2}\par
 \endgroup}
\makeatother

\def\tensor{\otimes}
\def\d{\partial}
\def\BE{\begin{equation}}   
\def\EE{\end{equation}}
\def\BA{\begin{array}}      
\def\EA{\end{array}}
\def\req#1{(\ref{#1})}

\def\cH{{\cal H}}
\def\cA{{\cal A}}
\def\cB{{\cal B}}
\def\cC{{\cal C}}

\def\cG{{\cal G}}
\def\cJ{{\cal J}}

\def\cL{{\cal L}}
\def\cM{{\cal M}}

\def\cQ{{\cal Q}}

\def\cT{{\cal T}}
\def\cV{{\cal V}}

\def\cW{{\cal W}}

\def\bG{\bar\cG}

\def\sj{{\,\ssf j\,}}

\def\sq{{i\over{\sqrt 2}}}
\def\sq1{{1\over{\sqrt 2}}}
\def\kalphapl{{\sqrt 2\alpha_+}}

\def\emt{energy-momentum tensor}
\def\bar{\overline}

\def\ap{\alpha_+}
\def\am{\alpha_-}

\def\half{{\textstyle{1\over2}}}
\def\eighth{{\textstyle{1\over8}}}
\def\threehalves{{\textstyle{3\over2}}}

\def\ket#1{\bigl|{#1}\bigr\rangle}
\def\nket#1{|{#1}\rangle}

\begin{document}
\hfuzz=2pt

\addtolength{\parskip}{2pt}

\vbox to2cm{\vfill}


\bigskip

\begin{center}
\Large\sc $s\ell(2)_{-4}$ WZW Model as an $N\!=\!4$-Supersymmetric Bosonic
String with $c\!=\!-2$ matter
\end{center}

\bigskip

\begin{center}
{\large A.~M.~Semikhatov}\\[2ex]
and\\[2ex]
{\large I.~Yu.~Tipunin}\\[4ex]
{\small\sl I.E.~Tamm Theory Division, P.~N.~Lebedev Physics Institute\\[-1pt]
Russian Academy of Sciences, 53 Leninski prosp., Moscow 117924, Russia}\\
\end{center}
\vskip4ex
\centerline{\sc abstract}
\vskip1.2ex
\centerline{\parbox{.9\hsize}{\addtolength\baselineskip{-2ex}
\noindent\small
We consider the $s\ell(2)$ current algebra at level $k=-4$ when the
$s\ell(2)$ BRST operator is nilpotent. We formulate a spectral sequence
converging to the cohomology of this BRST operator. At the second term of the
spectral sequence, we observe an $N\!=\!4$ algebra. This algebra is generated
in a $c=-2$ bosonic string whose BRST operator $\cQ_{\rm string}$ represents
the next term in the spectral sequence. We realize the cohomology of the
irreducible modules as $\cQ_{\rm string}$-primitives of the $N\!=\!4$
singular vectors and point out their relation to Lian--Zuckerman states
of $c=-2$ matter.  The relation between $\SL2_{-4}$ WZW model and $c=-2$
bosonic string is established both at the level of BRST cohomology and at the
level of an explicit operator construction.  The relation of the $N\!=\!4$
algebra to the known symmetries of matter+gravity systems is also
investigated.}}

\thispagestyle{empty}

\setcounter{page}{0}
\newpage
\section{Introduction \label{INTROD}}\lvm
In this paper we will consider the $\SL2$ current algebra at level $k=-4$.
The BRST operator is then nilpotent quantum-mechanically without the
introduction of an auxiliary set of currents, and the cohomology problem can
be posed for the algebra itself (with the appropriate ghosts) rather than for
a coset space.  The cosets of $\SL2$ for arbitrary $k$ have been considered
in a number of papers, e.g.\ in~\cite{[AGSY],[HY],[BMP],[HR]}.  Choosing
$k=-4$ allows us to use several specific tools, such as a spectral sequence
that converges to the $\SL2$ BRST cohomology and leads to uncovering an
underlying (level-1) $\N4$ superconformal algebra. The $sl(2)$ singular
vectors then take the form of singular vectors in an $\N4$ module, and these
can be used to produce Lian-Zuckerman states~\cite{[LZ]} in the corresponding
$c=-2$-matter + gravity theory.

The correspondence between the $\SL2$ WZW model and a matter+gravity theory,
which follows from the cohomological analysis, can also be arrived at in
rather explicit terms, by using a representation for $\SL2$ currents obtained
by inverting the hamiltonian reduction~\cite{[S-sing]}.  This representation
is built by tensoring a {\it Virasoro\/} Verma module with free-field modules
of the Liouville and ghost fields in such a way that the Hamiltonian
reduction maps it back into the Virasoro module, while the Liouville and
ghost fields correspond to those of a non-critical bosonic string. The matter
theory is chosen according to the $\SL2$ level $k$ and in our case of
$k\!=\!-4$, the $\SL2$ representation in question leads to an explicit
relation between the $\SL2_{k=-4}$ algebra and the bosonic string with $c=-2$
matter~\cite{[Distler]} (cf.\ the previous analysis~\cite{[MV]} of the $k=-3$
WZW model which has lead to $c=1$ matter).

The mappings between $\SL2$ and matter+gravity models can be
viewed in a more general setting of Universal string theory
\cite{[BV],[Fof],[IK]}, in which equivalences between theories with different
underlying symmetry algebras are established.  Two models of conformal field
theories are usually considered equivalent if they lead to identical physical
results, which can be taken to hold in (at least) three ways: as equivalence of
their BRST cohomologies, as isomorphism of the correlators, or as
`reducibility' of singular vectors of one of the underlying algebras to those
of the other algebra.  These three approaches to establishing isomorphisms
between different theories are heuristically equivalent. Indeed, correlators
(to be precise, conformal blocks) in conformal field theories are solutions
to differential equations derived from vanishing conditions of singular
vectors. Singular vectors are in turn related to BRST cohomology in the
following way. One observes that singular vectors in Verma modules are
BRST-exact, $\nket{{\rm sing}}=\cQ_0\nket\psi$, where $\nket\psi$ cannot be
BRST-exact; now in the (usually irreducible) module obtained by factoring with
respect to a submodule generated by $\nket{{\rm sing}}$, $\nket\psi$ becomes
BRST-closed and therefore represents a state in the cohomology.  Within the
BRST-cohomology approach, a useful tool for establishing isomorphism between
cohomology spaces is given by homotopy transformations. A homotopy
transformation is required to provide a splitting of the BRST-operator of a
given theory into a sum of two BRST-operators, one of which has trivial
cohomology and the other coincides with the BRST operator of another theory,
whose equivalence to the first theory is to be established. However, such an
explicit mapping between the respective BRST cohomology spaces is not always
easy to find; fortunately, the homotopy transformation approach is a special
case of a more powerful technique of spectral sequences: a homotopy
transformation can be viewed as such a lucky choice of grading that results
in a spectral sequence with only two terms. In general, spectral sequences
contain more terms and therefore have a broader applicability.  On the other
hand, they are less convenient, as compared with homotopy transformations,
for establishing {\it explicit\/} mappings between states in the cohomology.

In this paper, we will trace, both at the cohomological level and in terms of
explicit operator constructions, the relation between $\SL2_{-4}$ theory and
the $c\!=\!-2$ matter dressed with ghosts and Liouville fields. We will
introduce a spectral sequence on the $\SL2_{-4}$ BRST complex and use it in
combination with an explicit operator realization of the $\SL2$ currents.
The $\SL2$ representation that we use is not a free-field `bosonization', as
it involves as a building-block an arbitrary Virasoro Verma module which need
not be bosonized through free fields. As a result, it does not lead to an
`accidental' vanishing of singular vectors and therefore can be applied to
the description of the cohomology in irreducible modules `generated' by
singular vectors as explained above. We will thus systematically work with
singular vectors rather than with the corresponding cohomology elements.

It is a remarkable fact that a generating construction~\cite{[MFF]} (to be
referred to as MFF) is known for the $\SL2$ (in fact, $\SL{n}$) singular
vectors. Our representation for $\SL2$ currents can be made `compatible' with
the spectral sequence converging to the BRST cohomology, which will allow us
to map the MFF singular vectors into a class of singular vectors in a
representation of the $(\N4)_{k=1}$ algebra (the corresponding reformulation
of the whole MFF {\it generating formula\/} would then be a variation on a
similar construction for the $\N2$ algebra~\cite{[ST2]}).  These $\N4$
singular vectors give rise to the $c=-2$ Lian--Zuckerman states, thereby
providing a `Lie-algebraic' origin of the latter. Their ghost numbers, in
particular, can be derived from the embedding diagram of the $\SL2_{k=-4}$
Verma modules.  We will also consider relations of the $\N4$ algebra emerging
in the spectral sequence with other algebras known to be relevant in
non-critical bosonic strings~\cite{[GS3],[BLNW]}.

\medskip

In section~2, we introduce a spectral sequence for the $\SL2_{-4}$ BRST
operator and observe an $\N4$ algebra in its first term.
In section~3, we map the $\SL2_{-4}$ singular vectors into singular
vectors in an $\N4$ module; then the $\SL2_{-4}$ cohomology can be
considered entirely in $\N4$ terms. We also give here the
embedding diagram of the $\SL2_{k=-4}$ Verma modules, which will then
`project' onto the $c=28$ Virasoro Verma module embedding diagram, related to
the Lian--Zuckerman states.  In section~4, the use of the $\SL2$
representation from ref.~\cite{[S-sing]} leads us to identifying a $c=-2$
bosonic string `inside' the $\SL2_{-4}$ theory.
The formula for $\N4$ singular vectors then projects, on the one hand, into
singular vectors in the $c=28$ Verma module, and on the other hand, provides
us with a construction for a class of Lian--Zuckerman states in the $c=-2$
bosonic string.  In section~5, we finally discuss the relation of the
observed $\N4$ symmetry with the known symmetries~\cite{[GS3],[BLNW]} of
matter dressed with gravity.  The appearance of the $\N4$ algebra will be
interpreted as a result of fitting together two twisted $\N2$ algebras known
to exist when dressing a matter theory with ghosts and the
Liouville~\cite{[GS3]}.

\section{$s\ell(2)$ BRST operator, spectral sequence, and $N\!=\!4$
\label{BRSTOPER}}
\subsection{$s\ell(2)$ currents, ghosts, and the BRST operator}\lvm In this
subsection, we fix our notations and introduce the BRST complex, Weyl group
action, and the representations of $\SL2$ we are going to consider in this
paper.  To begin with, the $s\ell(2)$ current algebra operator
products are taken in the form
\BE\new\BA{rcl}
J^0(z)J^{\pm}(w)&=&\pm{J^{\pm}\over{z-w}}\\
J^{+}(z)J^{-}(w)&=&{-k/2\over{(z-w)^2}}-{J^0\over{z-w}}\\
J^0(z)J^0(w)&=&{k/2\over{(z-w)^2}}
\label{SL2}\EA\EE
We also introduce three ghost systems $\cB^{+}$, $\cC_+$; $\cB^0$, $\cC_0$,
and $\cB^{-}$, $\cC_-$ associated with the  $s\ell(2)$ generators $J^+$,
$J^0$ and $J^-$ respectively.  Together, these make up an algebra
\BE
\cA=s\ell(2)\oplus [\cB^+,\cC_+]\oplus [\cB^-,\cC_-]\oplus [\cB^0,\cC_0]
\label{A}
\EE
where $[\cB^+,\cC_+]$, \ $[\cB^-,\cC_-]$ and $[\cB^0,\cC_0]$ are
superalgebras spanned by the corresponding ghost systems.  The full
energy-momentum tensor, which will be denoted by $\cT_\cA$, is equal to the
sum of Sugawara and ghost energy-momentum tensors and for $k=-4$ reads
\BE
\cT_\cA=\half(-J^0J^0 + J^+J^- + J^-J^+ )
-\cB^+\d\cC_+ - \cB^-\d\cC_- - \cB^0\d\cC_0
\label{EMT}
\EE
All the $\cB$ ghosts thus have dimension 1.  In the formula \req{EMT} and
other similar equations below, the ghost monomials are normal-ordered with
respect to the $sl_2$-invariant ghosts vacua.

For $k=-4$, the algebra $\cA$ is made into a BRST complex by introducing the
BRST current according to the standard recipe \cite{[KSch]}
\BE
\cJ_\cA=\cC_+J^{+}+\cC_0J^0+\cC_-J^{-}-\cC_-\cC_0\cB^{-}-
\cC_0\cC_+\cB^{+}-\cC_-\cC_+\cB^0\,.
\label{BRST}
\EE
The corresponding BRST charge
\BE
\cQ_\cA=\oint{\cJ_\cA}
\label{QA}
\EE
is indeed nilpotent, $\cQ_\cA^2=0$, when $k$ is equal to minus two Coxeter
numbers, $k=-4$ \cite{[AGSY]} \footnote{ To avoid misunderstanding, let us
remind the reader that, in our normalization (which differs from that adopted
in refs.~\cite{[AGSY],[ISRA]}), `the other' critical value, at which the
universal enveloping algebra acquires an infinite-dimensional
center~\cite{[Frenkel]}, is $k=-2$.}.  In this case the BRST {\it current\/}
is also OPE-isotropic, $\cJ_\cA(z)\cJ_\cA(w)=0$.  In the following, $k$ will
be set equal to $-4$.

The \emt\ \req{EMT} turns out to be BRST-{\it exact\/}:
\BE
\cT_\cA(z)=[\cQ_\cA,\cG_\cA(z)]\,,\qquad
\cG_\cA=\cB^+J^--\cB^0J^0+\cB^-J^+\,.
\label{Texact}\EE
It follows then that all the states in the cohomology of $\cQ_\cA$ must have
vanishing dimension.  Another condition on the cohomology follows by
considering the currents
\BE
{\hat J}^{\pm,0}= [\cQ_\cA,\,B^{\pm,0}]\,,\qquad\left\{
\new\BA{rcl}
{\hat J}^+&=&J^+ - \cB^+\cC_0 + \cC_-\cB^0\,,\\
{\hat J}^0&=&J^0 + \cB^+\cC_+ - \cB^-\cC_-\,,\\
{\hat J}^-&=&J^- + \cB^-\cC_0 - \cC_+\cB^0\,\\
\EA\right.
\label{hatJ}\EE
that satisfy an $s\ell(2)$ algebra at level $k+4=0$ (hence, in particular,
${\hat J}^0$ is OPE-isotropic).  Since ${\hat J}^0$ is BRST-trivial, states
in the cohomology must have ${\hat J}^0$-spin equal to zero.

The current $\cJ_\cA$ itself is also BRST-exact,
\BE
\cJ_\cA=[\cQ_\cA,\,\cH_\cA]\,,\qquad
\cH_\cA\equiv \cB^+\cC_+ + \cB^-\cC_- + \cB^0\cC_0\,.
\label{hatJexact}
\EE
Note that the cohomology is naturally graded by zero mode of $\cH_\cA$.
As is well known \cite{[AGSY]}, cohomology of $\cQ_\cA$ is given by
$H_{\rm rel}^*\oplus(\cC_0)_0 H_{\rm rel}^*$ where $H_{\rm rel}^*$ denotes
the {\it relative\/} cohomology, which will be the only one we are going to
consider.

\smallskip

It will be useful to extend the action of the affine Weyl group $\tilde W$ on
the $\cB^{\pm}\cC_{\pm}$ ghosts by demanding that $\tilde W$ act on the $\hat
J$ currents in the same way as it acts on $J^{\pm,0}$, namely, for $\tilde
W\ni(s,\ell)$ where $s\in W$ is $+$ or $-$ and $\ell$ is an element of the
weight lattice,
\BE\new\BA{rcl}
(s,\ell)\,.\,{\hat J}^\alpha_n&=&{\hat J}^{s\alpha}_{n+\alpha\ell}\,,\\
(s,\ell)\,.\,{\hat J}^0_n&=& s\,{\hat J}^0_n\,.
\EA\label{W1}\EE
It follows then that
\BE\new\BA{rcl}
(s,\ell)\,.\,\cB^\alpha_n&=&s\,\cB^{s\alpha}_{n+\alpha\ell}\,,\\
(s,\ell)\,.\,(\cC_\alpha)_n&=&s\,(\cC_{s\alpha})_{n-\alpha\ell}\,.
\EA\label{W2}\EE

\smallskip

Now we specify a representation of the algebra \req{A}. Consider a
highest-weight representation of $s\ell(2)$ with an integral spin $j$:
\BE\new\BA{rcl}
J^0_0\ket{j}&=&j\ket{j}\,,\qquad
J^0_n\ket{j}~=~0,\quad n\geq1\\ J^+_n\ket{j}&=&0\,,\quad
n\geq0\\ J^-_n\ket{j}&=&0\,,\quad n\geq1\,\label{sl2highest}
\EA\EE
and tensor it with the corresponding ghost vacua into a `highest-weight' of
$\cA$:
\BE
\ket{j}_\cA=
\left\{\new\BA{ll}
\ket{j}\tensor\ket{j+1}_+ \tensor\ket0_0 \tensor \ket1_-
    &j\geq0\,,\\
\ket{j}\tensor\ket{0}_+ \tensor\ket0_0 \tensor \ket{-j}_-
    &j<0
\EA\right.
\label{jdressed}
\EE
The ghost vacua $\ket{0}_{\pm,0}$ are defined as follows. For a $bc$ system
of dimension $\lambda$, we define $\ket q$ by~\cite{[FMS]}
\BE
b_{\geq1-\lambda+q}\ket{q}=0\,,\qquad c_{\geq\lambda-q}\ket{q}=0
\label{ghostconditions}
\EE
(this is $sl_2$-invariant for $q=0$). This state has dimension
\BE
\Delta_\lambda(q)=\half q(q+1-2\lambda)\,,
\EE
the formula to be used when dressing $\SL2$ states with ghosts so as to
get dimension-0 states (as will be necessary for states in the cohomology).
All monomials in $b,c$ will always be assumed normal-ordered with respect to
the $sl_2$-invariant vacuum $\ket0$. Then, in particular
\BE
(bc)_0\ket{q}=-q\ket{q}\,,
\EE
which explains the choice of ghost states in \req{jdressed}, yielding the
vanishing ${\hat J}^0$-spin.  Now, to return to \req{ghostconditions}, the
ghost vacua in the formula \req{jdressed} for $j>0$, for instance, are such
that
\BE\new\BA{rclcrcl}
(\cC_+)_n\ket{j+1}_+&=&0,\quad n\geq-j\,,
&{}&{\cB^+}_n\ket{j+1}_+&=&0,\quad n\geq1+j\,,\\
(\cC_0)_n\ket{0}_0&=&0,\quad n\geq1\,,&{}&
{\cB^0}_n\ket{0}_0&=&0,\quad n\geq0\,,\\
(\cC_-)_n\ket{1}_-&=&0,\quad n\geq0\,,
&{}&{\cB^-}_n\ket{1}_-&=&0,\quad n\geq1
\label{sl2ghost}
\EA\EE
(recall that our ghost systems have dimensions $\lambda=1$).

The representation of the $\cA$ algebra is built over the
vacuum~\req{jdressed} by acting on it with the creation operators.

The ${\hat J}^0$-spin-zero vacuum state \req{jdressed} can be considered as a
representative of the only state in the cohomology of the {\it Verma\/}
module for integral $j$.
Simple considerations based on the `compensation' of the $\SL2$ spin $j$ by
the ghost contributions show that there is not even a vacuum in the
cohomology of the Verma module for half-integral $j$, and so we restrict
ourselves to the case of integral $j$. In what follows, we will look for the
cohomology of the irreducible modules obtained as factors of the Verma
module.

\subsection{Spectral sequence and $N\!=\!4$ algebra}\lvm
Since the \emt\ and the BRST current $\cJ_\cA$ are BRST-exact, one might
expect an underlying topological algebra in the system.  This is indeed the
case, and the topological algebra turns out to be a certain extension
\cite{[ISRA],[Kazama]} of the twisted $N\!=\!2$ algebra~\cite{[Ey],[W-top]}.
It differs from the twisted $N\!=\!2$ by new terms in the operator product
$\cG_\cA\cdot\cG_\cA$ (where $\cG_\cA$ is the superpartner to the \emt\ from
eq.~\req{Texact}):
\BE
\cG_\cA(z)\cG_\cA(w)={\cW\over z-w}\quad\hbox{where}\quad
\cW(z)=[\cQ_\cA,\cV(z)]\quad\hbox{and}\quad\cV=\cB^-\cB^+\cB^0\,.
\label{VW}\EE
$\cV$ and $\cW$ generate a commutative ideal in the extended topological
algebra, and the twisted $N\!=\!2$ algebra is a factor with respect to this
ideal. This extended topological algebra is not very useful to work with (at
least as compared to the true $\N2$), however, as we are going to show, this
is not needed, since the first step in evaluating the cohomology of $\cQ_\cA$
will effectively lead to factoring over the ideal generated by $\cV$ and
$\cW$.

\medskip

Observe that a filtration $F^i\cA$ exists on the BRST complex $(\cA,
\cQ_\cA)$ (i.e., the filtration is compatible with the action of the BRST
operator in the sense that $\cQ_\cA(F^i\cA)\subset F^i\cA$).  The filtration
can be described by first assigning the following {\it gradings\/} to our
fields:
\BE\new\BA{l}
\deg\cC_0=\deg\cB^0=\deg J^0=\deg \cB^-=\deg \cC_-=0\,,\\
\deg J^-=-\deg J^+=1\,,\\
-\deg \cB^+=\deg\cC_+=3\,.
\EA\label{degrees}\EE
The algebra $\cA$ is then decomposed into a direct sum of subspaces $G_l$
with definite degrees, and the filtration \ $\ldots\subset F^i\cA\subset
F^{i+1}\cA\subset F^{i+2}\cA\subset\ldots$ \ is defined by
$F^i\cA=\bigoplus_{l\geq i}G_l$.  Then we can split the BRST current into a
finite sum of terms with non-negative degrees:
\BE
\cJ=\ldots+0+\cJ^{(0)}+\cJ^{(1)}+\cJ^{(2)}+\cJ^{(3)}+0+\ldots
\label{Qdecomposition}\EE
where
\BE\new\BA{l}
\cJ^{(0)}=\cC_0{\hat J}^0\,,\\
\cJ^{(1)}=\cC_-J^-,\\
\cJ^{(2)}=\cC_+J^+,\\
\cJ^{(3)}=\cC_+\cC_-\cB^0.
\EA\label{SPSEQU}\EE

Since the degrees of all the non-vanishing terms in \req{Qdecomposition} are
non-negative, there exists a spectral sequence associated with this
filtration which converges to the cohomology of $\cQ_\cA$.  Observe that the
differentials $\cQ^{(0)}$, $\cQ^{(1)}$, $\cQ^{(2)}$, $\cQ^{(3)}$
corresponding to $\cJ^{(0)}$, $\cJ^{(1)}$, $\cJ^{(2)}$ $\cJ^{(3)}$
respectively are nilpotent separately and $\cQ^{(0)}$ anticommutes with
$\cQ^{(1)}$ and $\cQ^{(2)}$.  As can be seen from the form of
$\cQ^{(0)}=\oint\cJ^{(0)}$, it effectively imposes the constraint ${\hat
J}^0\sim0$.  Further, $\cQ^{(3)}$ is zero on the cohomology of $\cQ^{(0)}$,
hence the spectral sequence degenerates after the third term.  Cohomology of
the BRST operator \req{BRST} is therefore given by the cohomology of
$\cQ^{(2)}$ evaluated on the cohomology of $\cQ^{(1)}$ which is evaluated on
the cohomology of $\cQ^{(0)}$.

As follows from a careful reading of the last phrase, the first step in
the analysis of the spectral sequence consists therefore in restricting to
the cohomology of $\cQ^{(0)}$.  This will have an immediate effect on the
extended topological algebra referred to  in the beginning of this
subsection.  Namely, the fields $\cV$ and $\cW$ vanish on the cohomology of
$\cQ^{(0)}$ since they consist of terms which are either $\cQ^{(0)}$-exact or
not $\cQ^{(0)}$-closed. As a result, the extended topological algebra reduces
to the twisted $N\!=\!2$ algebra~\cite{[Ey],[W-top]}.  A useful choice
of representatives of the $N\!=\!2$ algebra generators is given by
\BE\new\BA{rcl}
\hat\cJ_\cA&=&\cC_-J^- + \cC_+J^+\,,\\
\hat\cG_\cA&=&\cB^-J^+ + \cB^+J^-\,,\\
\hat\cT_\cA&=&J^-J^+
+ \d\cB^+\cC_+ - 2\cB^-\d\cC_- - \d\cB^-\cC_- +
\half(J^0\cB^+\cC_+ - J^0\cB^-\cC_- + \d J^0) \,,\\
\hat\cH_\cA&=&\cB^+\cC_+ + \cB^-\cC_-\,.\EA\label{N2}\EE
These close to an $N\!=\!2$ algebra modulo $\cQ^{(0)}$-exact terms.

The structure of the cohomology of $\cQ^{(0)}$ can in fact be refined
considerably by noticing that it bears a representation of an $N\!=\!4$
algebra that extends the above $N\!=\!2$ algebra.  Representatives of the
$N\!=\!4$ generators can be chosen as
\BE\new\BA{rclcrcl}
\cT&=&\hat\cT_\cA\,,
&\qquad&\cG^1&=&\cC_+J^+\,, \\
J^+_{N=4} &=&\cC_-\cC_+\,,&{}&\cG^2&=&\cB^-J^+\,,\\
J^0_{N=4} &=&-\half\hat\cH_\cA\,,&{}&\bar\cG_1&=&\cB^+J^-\,,\\
J^-_{N=4} &=&\cB^-\cB^+\,,&{}&\bar\cG_2&=&\cC_-J^-\,.
\EA\label{GEN4ALGSL}\EE
Then it can be checked that the following $N\!=\!4$ OPEs~\cite{[Mats]} are
satisfied modulo $\cQ^{(0)}$-exact terms~\footnote{Greek superscripts and
subscripts $\alpha=0,\pm$ denote $s\ell(2)$ triplets, while Latin
superscripts and subscripts $a,b$ run over 1,2 and label $s\ell(2)$ doublet
and antidoublet representations.  The sigma matrices $\sigma^0,\sigma^+$ and
$\sigma^-$ are defined as:
$$\new\BA{rcl}
{\sigma^0=\left(\matrix{-\half&0\cr 0&\half\cr}\right)}&
{\sigma^+=\left(\matrix{0&0\cr 1&0\cr}\right)}&
{\sigma^-=\left(\matrix{0&-1\cr 0&0\cr}\right)}
\EA$$
Superscripts label rows and subscripts, columns.  The metric tensor
$\eta_{\alpha\beta}$ is:  $\half\eta_{00}=-\eta_{+-}=-\eta_{-+}=1$ and other
components are equal to zero.}:
\BE\new\BA{rclcrcl}
J_{N=4}^{\alpha}(z)\cG^a(w)&=&-{(\sigma^{\alpha })^a_b\cG^b\over z-w},&{}&
J_{N=4}^{\alpha}(z)\bar\cG_a(w)&
=&{\bar\cG_b(\sigma^{\alpha})_a^b\over z-w}\,,\\
\cT(z)\cG^a(w)&=&{a\cG^a(w)\over(z-w)^2}+{\d\cG^a\over z-w}\,,
&{}&
\cT(z)\bar\cG_a(w)&=&{(3-a)\bar\cG_a(w)\over(z-w)^2}+
{\d\bar\cG_a\over z-w}\\
\cT(z)J_{N=4}^{\alpha}(w)&=&\multicolumn{5}{l}{
{-\delta_{\alpha,0}\over(z-w)^3} +
{(1-\alpha)J_{N=4}^{\alpha}(w)\over(z-w)^2}+
{\d J_{N=4}^{\alpha}\over z-w}\,,\quad\alpha=+1,0,-1\,,}\\
\cG^a(z)\cG^b(w)&=&0,&{}&
\bar\cG_a(z)\bar\cG_b(w)&=&0,\\
\multicolumn{7}{l}{
\cG^a(z)\bar\cG_b(w)={2\,\delta^a_b\over(z-w)^3}-
{2(\sigma^{\alpha})^a_b\eta_{\alpha\beta} J_{N=4}^{\beta} (w)
\over(z-w)^2}+
{-(\sigma^{\alpha})^a_b\eta_{\alpha\beta}\d J_{N=4}^{\beta}
+\delta^a_b(\cT-\d J_{N=4}^0)\over z-w}}
\label{SUPERALGSL}\EA\EE
where $J_{N=4}^{+}$, $J_{N=4}^0$ and $J_{N=4}^-$ make up an $s\ell(2)$
algebra at level 1:
\BE\new\BA{rclcrcl}
 J_{N=4}^0 (z) J_{N=4}^{\pm} (w)&=&{ J_{N=4}^{\pm}\over z-w}
,&{}&  J_{N=4}^{+} (z) J_{N=4}^{-} (w)&=&
-{1\over(z-w)^2}-{2 J_{N=4}^0 \over z-w} ,\\
 J_{N=4}^0 (z) J_{N=4}^0 (w)&=&{1/2\over(z-w)^2}\,.
\EA\EE
This is in fact a {\it twisted\/} algebra, in particular $J_{N=4}^{\pm}$ have
dimensions $1\mp1$, so that the corresponding commutation relations read:
\BE
[(J_{N=4}^+)_m,\,(J_{N=4}^-)_n]=-\delta_{m+n,0}(m-1) - 2 (J_{N=4}^0)_{m+n}\,.
\EE

\medskip

The Weyl group action \req{W1}, \req{W2} carries over to the $\N4$ algebra.
Translations along the weights from the affine Weyl group act on the
generators trivially, while the reflection `$-$' acts as
\BE\new\BA{rclcrcl}
\cT&\mapsto&\cT\,,&\qquad&\cG^1&\mapsto&-\bar\cG_2\,, \\
J^+_{N=4} &\mapsto&-J^+_{N=4}\,,&{}&\cG^2&\mapsto&-\bar\cG_1\,,\\
J^0_{N=4} &\mapsto&J^0_{N=4}\,,&{}&\bar\cG_1&\mapsto&-\cG^2\,,\\
J^-_{N=4} &\mapsto&-J^-_{N=4}\,,&{}&\bar\cG_2&\mapsto&-\cG^1
\EA\label{N4Weyl}
\EE
where the minus signs in front of the fermionic generators
can be omitted without affecting the $\N4$ commutation
relations. Evaluating the transformation of $\cT$ when this $\N4$ generator
is represented as in \req{GEN4ALGSL}, we find, literally,
$\cT\mapsto\cT+2\d\hat J^0$ but this does not actually change $\cT$ as an
element in the cohomology of $\cQ^{(0)}$.

Note also that $N\!=\!4$ algebra admits, along with~\req{N4Weyl}, another
automorphism:
\BE\new\BA{rclcrcl}
\cT&\to&\cT-2\d J^0_{N=4}\,,&{}& \cG^1&\to&\cG^2,\\
J^+_{N=4}&\to&- J^-_{N=4}\,,&{}&\cG^2&\to&\cG^1\\ J^0_{N=4}&\to&-
J^0_{N=4}\,,&{}&\bar\cG_1&\to&\bar\cG_2\\ J^-_{N=4}&\to&-
J^+_{N=4}\,,&{}&\bar\cG_2&\to&\bar\cG_1 \EA\label{automorphism}\EE
which is induced in the construction \req{GEN4ALGSL} by interchanging the
ghosts as \BE\new\BA{rclcrcl}
\cB^+&\rightarrow&\cC_-\,,&{}&\cC_-&\rightarrow&\cB^+\,,\\
\cC_+&\rightarrow&\cB^-\,,&{}&\cB^-&\rightarrow&\cC_+\,.\\
\EA\label{pairs1}\EE

\subsection{Representations}\lvm
Now let us see what representation of the $\N4$ algebra is arrived at
starting with an $\SL2$ representation.  We consider the $\SL2$
highest-weight representations built on highest weights $\ket{j}$ with $j={}$
$j_+(r,s)$ or $j_-(r,s)$, labeled by two positive integers $r$ and $s$
via
\begin{eqnarray}
j_+(r,s)&=&\half(r-1)-\half(k+2)(s-1)\,,\qquad r,s\geq1\,,\label{jplus}\\
\noalign{\noindent and}
j_-(r,s)&=&-\half(r+1)+\half(k+2)s\,,\qquad r,s\geq1\,.\label{jminus}
\end{eqnarray}
Such states will be eigenstates of the $\N4$ current $J^0_{N=4}$:  we
evaluate $2(J^0_{N=4})_0$ on $\ket{j}_\cA$ as
\BE
2(J^0_{N=4})_0\ket{j}_\cA=
\left\{\new\BA{ll}
(j+2)\ket{j}_\cA\,,&j+2={r-1\over2}+s+1\geq2\,,\\
-j\ket{j}_\cA\,,&-j={r+1\over2}+s\geq2
\EA\right.
\EE
Therefore, fixing a $\sj\geq2$ that will be the eigenvalue of
$2(J^0_{N=4})^{\phantom{Y}}_0$ -- (twice) the {\it N=4 spin\/} -- we arrive
at {\it two\/} $\N4$ states defined in the following way: for each of these
states,
\BE\new\BA{rcllcrcll}
{\cL}^{\phantom{Y}}_n\ket{\sj,\pm}_{N=4}&=&0\,, &n\geq0\,, \\
(J^0_{N=4})^{\phantom{Y}}_n\ket{\sj,\pm}_{N=4}&=&0\,, &n \geq 1\,, &{\quad}&
   (J^+_{N=4})^{\phantom{Y}}_n\ket{\sj,\pm}_{N=4}&=&0\,, &n \geq -\sj+1\,,\\
2(J^0_{N=4})^{\phantom{Y}}_0\ket{\sj,\pm}_{N=4}&=&
\multicolumn{2}{l}{\sj\ket{\sj,\pm}_{N=4}\,,} &{}&
(J^-_{N=4})^{\phantom{Y}}_n\ket{\sj,\pm}_{N=4}&=&0\,, &n \geq \sj-1\,,
\EA\label{skewed0}\EE
while
\BE\new\BA{rcllcrcll}
(\cG^1)_n\ket{\sj,+}_{N=4}&=&0\,, &n\geq -\sj+1\,, &{\quad}&
                            (\cG^2)_n\ket{\sj,+}_{N=4}&=&0\,, &n\geq 0\,,\\
(\bar\cG_1)_n\ket{\sj,+}_{N=4}&=&0\,, &n\geq \sj-1\,,&{}&
  (\bar\cG_2)_n\ket{\sj,+}_{N=4}&=&0\,, &n\geq 0\,,
\EA
\label{skew4plus}\EE
whereas
\BE\new\BA{rcllcrcll}
(\cG^1)_n\ket{\sj,-}_{N=4}&=&0\,, &n\geq 0\,, &{\qquad}&
        (\cG^2)_n\ket{\sj,-}_{N=4}&=&0\,, &n\geq \sj-1\,,\\
(\bar\cG_1)_n\ket{\sj,-}_{N=4}&=&0\,, &n\geq 0\,,&{}&
        (\bar\cG_2)_n\ket{\sj,-}_{N=4}&=&0\,, &n\geq -\sj+1\,.
\EA
\label{skew4minus}\EE
Note that the two types of highest-weight conditions, eqs.~\req{skew4plus}
and~\req{skew4minus}, are related by the Weyl reflection \req{N4Weyl}
on the $\N4$ generators\footnote{Note similar `skewed' highest-weight
conditions in~\cite{[PT]}.}.

One can notice that the above `skewed' highest-weight states
$\ket{\sj,\pm}_{N=4}$ are related to $\N4$ highest weights that exist in the
series of spin-$\half$ $(\N4)_{k=1}$ representations labeled by conformal
dimensions~\cite{[ET]}. Namely, consider a highest-weight state
$\ket{\half,\Delta}_{N=4}$ satisfying the conditions
\BE\new\BA{l}
\cL_{\geq1}\ket{\half,\Delta}_{N=4}=
(J^{\pm,0}_{N=4})_{\geq1}\ket{\half,\Delta}_{N=4}=
(\cG^a)_{\geq1}\ket{\half,\Delta}_{N=4}=
(\bG_a)_{\geq1}\ket{\half,\Delta}_{N=4}=0\,,\\
(\cG^2)_0\ket{\half,\Delta}_{N=4}=(\bG_1)_0\ket{\half,\Delta}_{N=4}=0\,,\\
\cL_0\ket{\half,\Delta}_{N=4}=\Delta\ket{\half,\Delta}_{N=4}\,,\quad
(J^0_{N=4})_0\ket{\half,\Delta}_{N=4}=\half\,\ket{\half,\Delta}_{N=4}\,.
\label{EThighest}\EA\EE
(all the Verma modules built over any $\ket{\half,\Delta}_{N=4}$
except $\ket{\half,0}_{N=4}$ are not unitary;
The $\ket{\half,0}_{N=4}$ representation is called massless~\cite{[ET]}).

In ref.~\cite{[ET]}, singular vectors of $\N4$ algebra were arrived at by
restricting to a representation of the $s\ell(2)$ subalgebra of the $\N4$
algebra and noticing that singular vectors of the $s\ell(2)$ subalgebra are
singular vectors of $\N4$ algebra. For unitary representations, other
singular vectors are absent~\cite{[ET]}.  As we will see, more singular
vectors exist for non-unitary representations.  In our case we have a
realization of the $\N4$ algebra in which the $s\ell(2)$ subalgebra is
constructed out of two ghost pairs, and therefore all singular vectors of
ref.~\cite{[ET]} vanish.  Thus the $\N4$ representations `induced' from the
$\SL2_{-4}$ Verma module tensored with ghosts are related to $\N4$ Verma
modules after factorization of the latter with respect to singular vectors
from ref.~\cite{[ET]}:  namely, the `skewed' highest-weight states
characterized by~\req{skewed0}--\req{skew4minus}, and the highest-weights
\req{EThighest} are related via
\BE
\ket{\sj,+}=(\cG^1)_{-\sj+1}\ldots(\cG^1)_{-1}\ket{\half,\Delta}_{N=4}
\qquad\hbox{when $\Delta=-\half \sj(\sj-1)$}\,
\label{dressinjpl}\EE
and, similarly,
\BE
\ket{\half,\Delta}_{N=4}=(\cG^2)_{0}\ldots(\cG^2)_{\sj-2}\ket{\sj,-}
\qquad\hbox{when $\Delta=-\half \sj(\sj-3)-1$}\,
\label{dressfromjmin}
\EE
Heuristically, our states $\ket{\sj,\pm}$ are each a `half' of the simplest
singular (or co-singular) vector in the `proper' highest-weight module.
The factorization leads to fulfillment of several conditions, including
\BE\new\BA{l}
\left((J^-_{N=4})_0\right)^2\ket{\half,\Delta}_{N=4}=0\,,\qquad
(\cG^2)_0(J^-_{N=4})_0\ket{\half,\Delta}_{N=4}=0\,,\qquad
(\bG_1)_0(J^-_{N=4})_0\ket{\half,\Delta}_{N=4}=0\,,\\
\left((J^+_{N=4})_0\right)^2\ket{\half,\Delta}_{N=4}=0\,,\qquad
(\cG^1)_0(J^+_{N=4})_0\ket{\half,\Delta}_{N=4}=0\,,\qquad
(\bG_2)_0(J^+_{N=4})_0\ket{\half,\Delta}_{N=4}=0\,.
\label{svectcond}\EA\EE
These vanishing conditions will be used in the next section.

\medskip

The states $\ket{\half,\Delta}_{N=4}$ (except
$\ket{\half,0}_{N=4}=\ket{-1}_\cA$) cannot be represented in intrinsic terms
of the algebra $\cA$, i.e.\ in terms of the $\SL2_{-4}$ algebra and ghosts.
Constructing the $\ket{\half,\Delta}_{N=4}$ states requires identifying in
the $\SL2$ representation a matter sector (the result of Hamiltonian
reduction) and the `complementary' ghost sectors, as will be shown in
section~4.

\medskip

To return to the spectral sequence, observe now that the differentials
$\cQ^{(1)}$ and $\cQ^{(2)}$ from \req{SPSEQU} are given by zero modes of the
$N\!=\!4$ generators $\bar\cG_2$ and $\cG^1$ respectively. Therefore the
cohomology of $\cQ_\cA$ \req{QA} will be given in the $\N4$ terms as the
cohomology of $(\cG^1)_0$ evaluated on the cohomology of $(\bar\cG_2)_0$.
Recall further that, while there is only a vacuum in the cohomology of the
Verma module, factoring with respect to submodules generated by singular
vectors gives rise to non-trivial cohomology; the standard
`cohomology-generating' mechanism relies on the fact that the
$s\ell(2)$-singular vectors are (upon dressing with ghosts appropriately)
BRST-trivial, and therefore their BRST-primitives become cohomological states
in the module where singular vectors vanish.  In our case, combining these
considerations with the existence of the spectral sequence, we will be able
to give a more detailed structure of the cohomology. In the next section, we
will consider how the cohomology of $\SL2$ is generated from the state
$\ket{j}_\cA$ by the $\N4$ algebra operators.  It looks plausible that the
cohomology of \label{hypothesis} $\cQ^{(0)}=\oint\cC_0{\hat J}^0$ is
generated precisely by the currents of the $N\!=\!4$ algebra.

\section{MFF vectors and cohomology}\lvm
In this section, we begin with the MFF singular states and then discuss how,
upon an appropriate dressing with ghosts, they rewrite as $\N4$ singular
states, and the further analysis can be carried out in terms of the $\N4$
algebra.

\subsection{MFF vectors and Verma modules at $k=-4$}\lvm
Consider singular states in the $\SL2$ Verma module built on the
highest-weight state $\ket{j}$. They are given by the MFF construction
\cite{[MFF]} and are labeled by two positive integers $r$ and $s$.  For
$j=j_+(r,s)$ (see~\req{jplus}) one has
\BE\new\BA{rcl}\ket{{\rm MFF}\{r,s\},-}&=&
(J^-_0)^{r+(s-1)(k+2)}(J^+_{-1})^{r+(s-2)(k+2)}(J^-_0)^{r+(s-3)(k+2)}
\ldots\\ {}&{}&{}\times (J^+_{-1})^{r-(s-2)(k+2)}
(J^-_0)^{r-(s-1)(k+2)}\ket{j_+(r,s)}\EA\label{mff}\EE
The MFF states $\ket{{\rm MFF}\{r,s\}}$ are annihilated by the same set of
annihilation operators as the highest-weight $\ket{j}$ (see~\req{sl2highest})
but have different spin and dimension (which for $k=-4$ are equal to $j-r$
and $-\half(j-r)(j-r+1)$ respectively).

Singular states determine the pattern of Verma module embeddings. In our case
of $k=-4$, every Verma module contains only a finite number of submodules
(corresponding to singular vectors) but can itself be embedded into an
infinite number of other modules ({\it co\/}singular vectors). Several
lower-$j$ embeddings that correspond to singular vectors \req{mff} are shown
here:
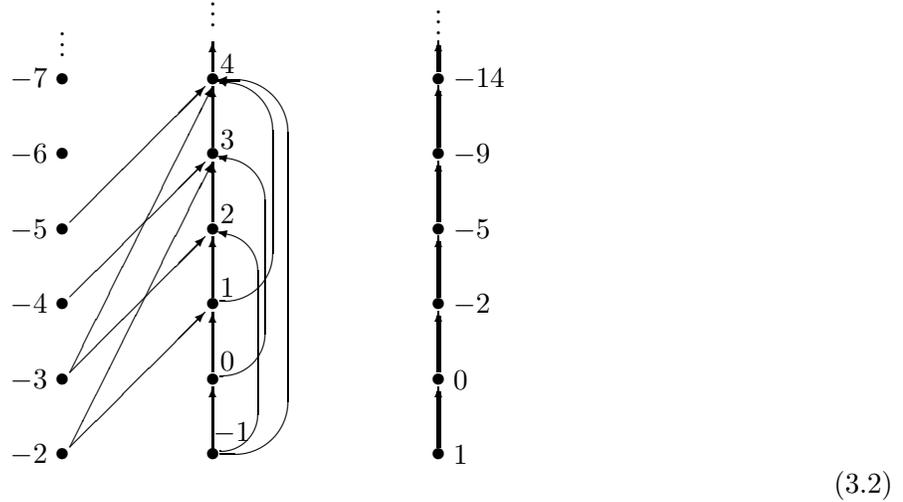
\begin{equation}
{
\unitlength=1.00mm
\begin{picture}(75.00,70.00)(40.00,05.00)
\put(70.00,10.00){\vector(0,1){9.00}}
\put(70.00,21.00){\vector(0,1){8.00}}
\put(70.00,31.00){\vector(0,1){8.00}}
\put(70.00,41.00){\vector(0,1){8.00}}
\put(70.00,51.00){\vector(0,1){8.00}}
\put(70.00,61.00){\vector(0,1){4.00}}
\put(69.50,67.00){$\vdots$}
\put(70.00,10.00){\makebox(0,0)[cc]{$\bullet$}}
\put(70.00,20.00){\makebox(0,0)[cc]{$\bullet$}}
\put(70.00,30.00){\makebox(0,0)[cc]{$\bullet$}}
\put(70.00,40.00){\makebox(0,0)[cc]{$\bullet$}}
\put(70.00,50.00){\makebox(0,0)[cc]{$\bullet$}}
\put(70.05,13.00){\makebox(0,0)[lc]{$-1$}}
\put(71.00,22.50){\makebox(0,0)[lc]{$0$}}
\put(71.00,32.30){\makebox(0,0)[lc]{$1$}}
\put(71.00,42.00){\makebox(0,0)[lc]{$2$}}
\put(71.00,52.00){\makebox(0,0)[lc]{$3$}}
\put(71.00,62.00){\makebox(0,0)[lc]{$4$}}
\put(51.00,11.00){\vector(1,1){18.00}}
\put(51.00,11.00){\vector(1,2){19.00}}
\put(51.00,21.00){\vector(1,1){18.00}}
\put(51.00,21.00){\vector(1,2){19.00}}
\put(51.00,31.00){\vector(1,1){18.00}}
\put(51.00,41.00){\vector(1,1){18.00}}
\put(70.00,60.00){\makebox(0,0)[cc]{$\bullet$}}
\put(50.00,20.00){\makebox(0,0)[cc]{$\bullet$}}
\put(50.00,30.00){\makebox(0,0)[cc]{$\bullet$}}
\put(50.00,40.00){\makebox(0,0)[cc]{$\bullet$}}
\put(50.00,50.00){\makebox(0,0)[cc]{$\bullet$}}
\put(50.00,60.00){\makebox(0,0)[cc]{$\bullet$}}
\put(50.00,10.00){\makebox(0,0)[cc]{$\bullet$}}
\put(48.00,10.00){\makebox(0,0)[rc]{$-2$}}
\put(48.00,20.00){\makebox(0,0)[rc]{$-3$}}
\put(48.00,30.00){\makebox(0,0)[rc]{$-4$}}
\put(48.00,40.00){\makebox(0,0)[rc]{$-5$}}
\put(48.00,50.00){\makebox(0,0)[rc]{$-6$}}
\put(48.00,60.00){\makebox(0,0)[rc]{$-7$}}
\put(49.50,63.00){$\vdots$}
\put(71.00,25.00){\oval(10.00,29.00)[r]}
        \put(73.55,39.03){\vector(-4,1){3}}
\put(71.00,35.00){\oval(12.00,29.00)[r]}
        \put(73.55,49.05){\vector(-4,1){3}}
\put(71.00,45.00){\oval(14.00,29.00)[r]}
        \put(73.55,59.08){\vector(-4,1){3}}
\put(71.00,35.00){\oval(18.00,49.80)[r]}
        \put(73.55,59.80){\vector(-4,0){3}}
\put(100.00,10.00){\makebox(0,0)[cc]{$\bullet$}}
        \put(102.00,10.00){\makebox(0,0)[lc]{$1$}}
\put(99.80,11.00){\line(0,1){7.40}}
\put(100.20,11.00){\line(0,1){7.40}}
\put(100.00,18.10){\vector(0,1){1}}
\put(100.00,20.00){\makebox(0,0)[cc]{$\bullet$}}
        \put(102.00,20.00){\makebox(0,0)[lc]{$0$}}
\put(99.80,21.00){\line(0,1){7.40}}
\put(100.20,21.00){\line(0,1){7.40}}
\put(100.00,28.10){\vector(0,1){1}}
\put(100.00,30.00){\makebox(0,0)[cc]{$\bullet$}}
        \put(102.00,30.00){\makebox(0,0)[lc]{$-2$}}
\put(99.80,31.00){\line(0,1){7.40}}
\put(100.20,31.00){\line(0,1){7.40}}
\put(100.00,38.10){\vector(0,1){1}}
\put(100.00,40.00){\makebox(0,0)[cc]{$\bullet$}}
        \put(102.00,40.00){\makebox(0,0)[lc]{$-5$}}
\put(99.80,41.00){\line(0,1){7.40}}
\put(100.20,41.00){\line(0,1){7.40}}
\put(100.00,48.10){\vector(0,1){1}}
\put(100.00,50.00){\makebox(0,0)[cc]{$\bullet$}}
        \put(102.00,50.00){\makebox(0,0)[lc]{$-9$}}
\put(99.80,51.00){\line(0,1){7.40}}
\put(100.20,51.00){\line(0,1){7.40}}
\put(100.00,58.10){\vector(0,1){1}}
\put(100.00,60.00){\makebox(0,0)[cc]{$\bullet$}}
        \put(102.00,60.00){\makebox(0,0)[lc]{$-14$}}
\put(99.80,61.00){\line(0,1){3.40}}
\put(100.20,61.00){\line(0,1){3.40}}
\put(100.00,64.10){\vector(0,1){1}}
\put(99.50,66.00){$\vdots$}
\end{picture}
}\label{picture}
\end{equation}
The numbers give the values of $j$; it should not be forgotten that there is
an infinite number of arrows going out of any dot to the `higher' ones. The
right column with double arrows represents {\it Virasoro\/} Verma modules
obtained via Hamiltonian reduction (the numbers give dimensions).  The arrows
are drawn in the direction of {\it embeddings\/}.  There are precisely $j+1$
arrows entering a dot labelled by spin $j>0$.  This pattern is determined by
the fact that, for integral $j$ and negative integral $k$, there exist
several ways to represent the spin $j$ as $j_+(r,s)$ with positive integral
$r$ and $s$.

A `dual' version of this embedding diagram exists for $j=j_-(r,s)$
(see~\req{jminus}), based on the MFF vectors given by a construction similar
to \req{mff}.  The corresponding counterpart of \req{mff} reads
\BE\new\BA{rcl}\ket{{\rm MFF}\{r,s\},+}&=&
(J^+_{-1})^{r+(s-1)(k+2)}(J^-_0)^{r+(s-2)(k+2)}(J^+_{-1})^{r+(s-3)(k+2)}
\ldots\\ {}&{}&{}\times (J^-_0)^{r-(s-2)(k+2)}
(J^+_{-1})^{r-(s-1)(k+2)}\ket{j_-(r,s)}\EA\label{mffneg}\EE
(with $k\to4$).
In what follows, we will mainly give explicit expressions for constructions
related to the MFF vectors $\ket{{\rm MFF}\{r,s\},-}$, denoting them simply
as $\ket{{\rm MFF}\{r,s\}}$.

\subsection{From $\SL2_{-4}$ to $N\!=\!4$ singular vectors}\lvm
By dressing with ghosts, the $\SL2$ singular vectors can be made into
singular vectors in the $\N4$ representation considered above. Thus the
BRST-primitives that represent the cohomology in the corresponding
irreducible models, will be given by vectors in the $\N4$ representation;
therefore the cohomology of $\cQ_\cA$ is concentrated
in the $\N4$ term of the spectral sequence.  Now that we have an (almost)
explicit formula for the MFF singular vectors, it is interesting to see to
what extent it can be carried over to the $\N4$ algebra that arises in the
spectral sequence.  Taking an MFF state and tensoring it with the ghost vacua
from section~2, as
\BE
\ket{{\rm MFF}\{r,s\}}\tensor\ket{j+1}_+ \tensor\ket0_0 \tensor \ket1_-
\EE
we observe that this can be dressed with ghost modes so as to produce a state
with zero ${\hat J}^0$-spin: for $j>0$ (with the formula \req{mff} valid for
$\SL2$ singular vectors), the dressed states would read
\BE
\ket{{\rm MFF}\{r,s\}}_\cA=
\Biggl\{\!\!\new\BA{ll}
\!\ket{{\rm MFF}\{r,s\}}\tensor
\cB^+_{j-r+1}\ldots\cB^+_{j-1}\cB^+_{j}\ket{j+1}_+
\tensor\ket0_0 \tensor\ket1_-,& r\leq j\!+\!1 \\
\!\ket{{\rm MFF}\{r,s\}}\!\tensor\!
\cB^+_{0}\ldots\cB^+_{j-1}\cB^+_{j}\ket{j+1}_+
\!\tensor\!\ket0_0\!\tensor\!(\cC_-)_{j-r+1}\ldots(\cC_-)_{-1}\ket1_-,&
r>j\!+\!1
\EA
\label{mffdressed}\EE

The states thus obtained turn out to be $\cQ_\cA$-closed and, moreover,
$\cQ_\cA$-exact.  Then, any state
$\ket*$ such that
\BE
\ket{{\rm MFF}\{r,s\}}_\cA = \cQ_\cA\ket{{*}}
\label{MFFexact}
\EE
would be a representative in the $\cQ_\cA$-cohomology of the irreducible
module  obtained by factorization of the Verma module over the null vector
$\ket{\rm MFF\{r,s\}}$.

These cohomology elements occur in a particular term of the spectral sequence
associated with \req{SPSEQU}, namely as states in the representation of the
$\N4$ algebra from the previous section.  They can indeed be constructed by
acting with modes of the $N\!=\!4$ generators~\req{GEN4ALGSL} on the
vacuum~$\ket{\sj,+}_{N=4}$ (similarly, for $j<0$, the corresponding MFF
vectors \req{mffneg} can be dressed with ghosts in such a way that would
allow rewriting them as $\N4$ singular vectors built on the vacuum
$\ket{\sj,-}$).  For definiteness, we will consider explicitly the $\N4$
singular vectors built on the $\ket{\sj,+}$ vacua.

Consider first the $r1$ MFF states. They are built on the $\SL2$
highest-weight state of spin $j=\half(r-1)$ and can thus be written as
$\ket{{\rm MFF}\{2j+1,1\}}$~\footnote{more precisely, $\ket{{\rm
MFF}\{2j+1,1\},-}$; the `$+$'-counterpart reads $\ket{{\rm
MFF}\{k+1-2j,1\},+}$.}.  When dressed with the ghosts, the
states $\ket{{\rm MFF}\{2j+1,1\}}_\cA$ are identically
rewritten as elements of the $(\N4)_1$ representation, i.e.\ generated from
the vacuum by the action of the $\N4$ generators:
\BE
\ket{{\rm MFF}\{2\sj-1,1\}}_\cA=
(\bar\cG_2)_{-\sj+2}(\bar\cG_2)_{-\sj+3}\ldots(\bar\cG_2)_{-1}\,
(\bar\cG_1)_0\ldots(\bar\cG_1)_{\sj-2}\ket{\sj,+}_{N=4}
\label{mffn4}
\EE

Moreover, these states being $(\bG_2)_0$-exact, the corresponding
BRST-primitive state $\ket{*}$ in~\req{MFFexact} can be obtained by pulling
out the BRST operator $\cQ^{(1)}=(\bG_2)_0$. We then arrive at the following
representation for the primitives in terms of $N\!=\!4$ algebra generators
acting on the vacuum:
\BE
\ket{*}=(J^0_{N=4})_{-\sj+2}\,(\bar\cG_2)_{-\sj+3}
\ldots(\bar\cG_2)_{-1}\,
(\bar\cG_1)_0\ldots(\bar\cG_1)_{\sj-2}\ket{\sj,+}_{N=4}
\label{primitiveN4}
\EE
(as compared with \req{mffn4}, the leftmost mode $(\bar\cG_2)_m$ gets
replaced by the same mode of $J^0_{N=4}$).  This expression thus gives an
element in cohomology upon factorization over the submodule generated by the
MFF vector.

More generally, consider an arbitrary MFF vector for $k=-4$ with the only
condition that all the powers in the MFF formula \req{mff} be non-negative.
Among the $\ket{{\rm MFF},-}$-vectors these are $\ket{{\rm MFF}\{j+l,s\}}$
for $1\leq l\leq j+1$.  To elucidate their construction in terms of the $\N4$
algebra generators, we write them down together with the corresponding MFF
formula:
\BE
\new\BA{l}
\new\BA{lcl}
\ket{{\rm MFF}\{{2\sj-s-1},s\},-}_\cA=&{}&\ket{{\rm MFF}\{{2\sj-s-1},s\}}=\\
\quad(\bar\cG_2)_{-\sj+s}\ldots(\bar\cG_2)_{-1}\,
(\bar\cG_1)_0\ldots(\bar\cG_1)_{\sj-s}&&\quad(J^-_0)^{2\sj-2s+1}\\
\qquad(\cG^1)_{-\sj+s-1}\ldots(\cG^1)_{-1}\,
(\cG^2)_{0}\ldots(\cG^2)_{-\sj-s+1}\,&{}&\qquad(J^+_{-1})^{2\sj-2s+3}\\
\qquad\qquad\qquad\qquad\qquad\vdots&{}&\qquad\qquad\vdots\\
\qquad\qquad(\bar\cG_2)_{-\sj+4}\ldots(\bar\cG_2)_{-1}\,
(\bar\cG_1)_0\ldots(\bar\cG_1)_{\sj-4}&{}&\qquad\qquad(J^-_0)^{2\sj-7}\\
\qquad\qquad\quad(\cG^1)_{-\sj+3}\ldots(\cG^1)_{-1}\,
(\cG^2)_{0}\ldots(\cG^2)_{\sj-3}\,&{}&\qquad\qquad\quad(J^+_{-1})^{2\sj-5}\\
\qquad\qquad\qquad(\bar\cG_2)_{-\sj+2}\ldots(\bar\cG_2)_{-1}\,
(\bar\cG_1)_0\ldots(\bar\cG_1)_{\sj-2}\ket{\sj,+}_{N=4}&{}&
\qquad\qquad\qquad(J^-_0)^{2\sj-3}\ket{\sj\!-\!2}
\EA
\EA
\label{generalmffn4}
\EE
The $\N4$ spin (the eigenvalue of $2(J^0_{N=4})_0$) of this state is equal to
$l=\sj-2s+1$.
Every MFF factor $(J^{\pm}_{-1,0})^m$ corresponds to a group of $m$
factors given by modes of the $\N4$ generators.  The groups corresponding to
$(J^-_{0})^m$ consist of $\half(m+1)$ generators $\bar\cG_1$, their modes
ranging from $\half(m-1)$ to $0$ (recall that $m$ is always odd).  In
addition, the same group contains a product of modes of $\bar\cG_2$, from
$(\bar\cG_2)_{-1}$ to $(\bar\cG_2)_{{-m+1\over2}}$.  Thus, when passing the
zero mode, the $\N4$ generators inside one group get replaced according to
the action of the automorphism \req{automorphism} (however the left subgroup
is one element shorter).  To obtain the structure of the groups of $\N4$
factors corresponding to $(J^+_{-1})^m$, one should drop the leftmost and the
rightmost modes in the right neighbouring group (the one corresponding to
$(J^-_{0})^{m+2}$) and then act on the remaining modes with the Weyl
reflection \req{N4Weyl}.

\medskip

To check that the states constructed {\it are\/} singular vectors in the
$\N4$ module, consider the state obtained by acting on $\ket{\sj,+}_{N=4}$
with only the first (counting from the right) group in~\req{generalmffn4}
(the case $s=1$),
\BE
(\bar\cG_2)_{-\sj+2}\ldots(\bar\cG_2)_{-1}\,
(\bar\cG_1)_0\ldots(\bar\cG_1)_{\sj-2}\ket{\sj,+}_{N=4}\,.
\label{discuss}\EE
The eigenvalue of $2(J^0_{N=4})^{\phantom{Y}}_0$ on~\req{discuss}
is given by $\sj-({\rm\#\ of}\ \bar\cG_1)
+({\rm\#\ of}\ \bar\cG_2)=\sj-1$. The state~\req{discuss} is in fact a
$\ket{\sj-1,-}_{N=4}$. To see this, note first of all that it is annihilated
by $(\bar\cG_1)_{\geq0}$, since the modes $(\bar\cG_1)_{\geq\sj-1}$ used to
annihilate $\ket{\sj,+}_{N=4}$ while the remaining modes
$\cG_{0\leq n\leq\sj-2}$ will square to zero.
Similarly,~\req{discuss}
is annihilated by $(\bar\cG_2)_{\geq-\sj+2}$. This gives
a half of the highest-weight conditions \req{skew4minus} for spin $\sj-1$.
The other half can be deduced as follows.

Let us evaluate the action of $\cG^1_n$ on~\req{discuss} for $n\geq0$.  When
being plugged to the right, $\cG^1_n$ can hit one of the $\bar\cG_2$ or one
of the $\bar\cG_1$ modes.  Consider first
\BE{[}\,\cG^1_n,\,(\bG_2)_{-\sj+2}\ldots(\bG_2)_{-1}{]}\,
(\bG_1)_0\ldots(\bG_1)_{\sj-2}\ket{\sj,+}_{N=4}\,.
\label{work1}\EE
Using the $\N4$ commutation relations that follow from \req{SUPERALGSL}
(the brackets $[~,~]$ always mean the {\it super\/}commutator)
\BE\new\BA{rcl}
{[}\cG^a_n\,,(\bar\cG_b)_m{]}&=&\delta^a_b(n+a-1)(n+a-2)\delta_{m+n,0}
+(m-n+3-a-b)(\sigma^\alpha)^a_b\eta_{\alpha\beta}
 (J_{N=4}^\beta)_{m+n}^{\phantom{Y}}\\
&{}&{}+\delta^a_b\cL_{m+n}+(m+n+1)\delta^a_b(J_{N=4}^0)_{m+n}^{\phantom{Y}}\,,
\EA\EE
we see that $[\cG^1_n\,,(\bar\cG_2)_m]=(m-n)(J_{N=4}^+)_{m+n}^{\phantom{Y}}$
and the resulting $(J_{N=4}^+)_{m+n}$ can be moved to the right until it
meets $(\bar\cG_1)_0$. Commuting $(J_{N=4}^+)_{m+n}$ with the modes
$(\bar\cG_1)_r$ will produce $(\bar\cG_2)_{n+m+r}$.  Of these,
$(\bar\cG_2)_{p}$ with $p\geq0$ will annihilate the state
$\ket{\sj,+}_{N=4}$, while those with $-\sj+2\leq p\leq-1$ will square to
zero due to the presence of the same mode among the
$(\bar\cG_2)_{-\sj+2}\ldots(\bar\cG_2)_{-1}$ unless this mode has been
`spent' in the commutator $[\cG^1_n\,,(\bar\cG_2)_m]$; however, that would
never happen for $n>0$, and thus the result of commuting $\cG^1_n$ through
the $(\bar\cG_2)_{-\sj+2}\ldots(\bar\cG_2)_{-1}$ is
effectively zero for $n>0$.  When
$n=0$, however, the mode $(\bG_2)_m$ will be restored when commuting
$(J_{N=4}^+)_m$ with $(\bG_1)_0$, which gives
\BE\new\BA{l}
{[}\,\cG^1_0,\,(\bG_2)_{-\sj+2}\ldots(\bG_2)_{-1}{]}\,
        (\bG_1)_0\ldots(\bG_1)_{\sj-2}\ket{\sj,+}_{N=4}\\
{}\qquad{}=\sum_{m=-\sj+2}^{-1}m(-1)^{\sj+m}
(\bG_2)_{-\sj+2}\ldots~\Bigl/\!\!\!\!\!\!(\bG_2)_m\ldots(\bG_2)_{-1}\cdot
(\bG_2)_m\,(\bG_1)_1\ldots(\bG_1)_{\sj-2}\ket{\sj,+}_{N=4}\\
{}\qquad{}=\half(-1)^{\sj}(\sj-1)(\sj-2)
(\bG_2)_{-\sj+2}\ldots(\bG_2)_{-1}\cdot
(\bG_1)_1\ldots(\bG_1)_{\sj-2}\ket{\sj,+}_{N=4}
\EA\label{group1}\EE

It remains to see how $\cG^1_n$ commutes with
$(\bar\cG_1)_0\ldots(\bar\cG_1)_{\sj-2}$, namely to evaluate
\BE
(-1)^{\sj}(\bG_2)_{-\sj+2}\ldots(\bG_2)_{-1}\,
        {[}\,\cG^1_n,\,(\bG_1)_0\ldots(\bG_1)_{\sj-2}{]}\,\ket{\sj,+}_{N=4}
\label{group2}\EE
Here, in accordance with \req{SUPERALGSL}
\BE
{[}\,\cG^1_n\,,(\bar\cG_1)_m{]}=n(n-1)\delta_{m+n,0}
+2n(J_{N=4}^0)_{m+n}^{\phantom{Y}} + \cL_{m+n}\,,
\EE
and for $n=0$ we find $[\cG^1_0\,,(\bar\cG_1)_0]=\cL_0$; plugging $\cL_0$ to
the right amounts to adding up the mode numbers of
$(\bG_1)_1\ldots(\bG_1)_{\sj-2}$ as
$$
\sum_{r=1}^{\sj-2}(-r)=-\half(\sj-1)(\sj-2)\,,
$$
and the resulting term will precisely cancel~\req{group1}. All other
commutators in \req{group2} for $n=0$ give a vanishing contribution.  For
$n\geq1$, \req{group2} vanishes altogether.

Analyzing similarly the action of the modes \ $(J^+_{N=4})_{n\geq-\sj+2}$, \
$(\cG^2)_{n\geq\sj}$, and $(J^-_{N=4})_{n\geq\sj}$ on~\req{discuss}, we also
find that these are annihilators.  The necessary highest-weight relations
(see~\req{skew4minus}) require in addition two more vanishing conditions,
namely those for $(J^-_{N=4})_{\sj-1}$ and $(\cG^2)_{\sj-1}$. These are
satisfied by virtue of the relations~\req{svectcond}.

We thus conclude that the state
$(\bar\cG_2)_{-\sj+2}\ldots(\bar\cG_2)_{-1}\,
(\bar\cG_1)_0\ldots(\bar\cG_1)_{\sj-2}\ket{\sj,+}_{N=4}$ is indeed
proportional to $\ket{\sj-1,-}_{N=4}$.  Then the second from the right group
in~\req{generalmffn4} maps the resulting vector into $\ket{\sj-2,+}_{N=4}$
etc., which can be shown either directly or simply by noticing that the $\N4$
generators in the adjacent groups are related by the Weyl reflection
\req{N4Weyl} \footnote{and the relations~\req{svectcond} are Weyl-symmetric
as well.} (and the groups become shorter as moving from right to left,
according to how the $J^0_{N=4}$-spin decreases). The
vectors~\req{generalmffn4} are therefore singular in our $(\N4)_1$ module.

To compare with the formulation of ref.~\cite{[ET]}, we notice that a
singular vector in the module built over the `proper' highest
weight~\req{EThighest} can be constructed by means of the following
procedure. One dresses the state $\ket{\half,\Delta}_{N=4}$ \req{EThighest}
with $\Delta=-\half \sj(\sj-1)$ using the formula \req{dressinjpl} and
obtains a $\ket{\sj,+}$; then one builds over $\ket{\sj,+}$ the singular
vector \req{generalmffn4} and obtains $\ket{{\rm
MFF}\{{2\sj-s-1},s\},-}_\cA$.  And finally dressing it as in
\req{dressfromjmin} one obtains a new highest weight obeying~\req{EThighest}
with a new dimension
$\Delta=-\half(\sj-2s)(\sj-2s-1)$.
Therefore,
non-unitary $(\N4)_1$ modules are embedded into one another according to the
pattern described above, with the massless representation ($\Delta=0$) being
embedded into all the other.

\medskip

The $\cQ_\cA$-primitives of the singular
states thus constructed are given in the same
way as in \req{primitiveN4}, hence follow representatives in the cohomology
of the irreducible modules.

\medskip

Note that once the Weyl group action on the $\N4$ generators has been
identified, the mechanism behind the above derivation is very similar to the
one underlying the MFF construction. The analogies with the $\SL2$ MFF
formula are rather straightforward, however we have not carried out an
`analytic' continuation of the construction~\req{generalmffn4} off the
positive integer points. Recall that the power of the MFF construction is
that it can be given meaning for {\it all\/} values of the parameters, when
the exponents in~\req{mff} are complex numbers.  The $\N4$ analogue of the
`continued' formula would require introducing operators with non-integer
modding (which is the easy part of the construction) and replacing the
products of modes with intertwiners (which is somewhat more involved). We
actually expect a close analogy with the case of (twisted) $\N2$ algebra, for
which the `analytically' (in fact, {\it algebraically\/}) continued
construction can indeed be built~\cite{[ST2]}.  Yet we have not tried to
extend, in $\N4$ terms, the formula \req{generalmffn4} to the case of $r$ and
$s$ being arbitrary positive integers.  We have only checked in a number of
lower-level cases that the MFF singular vectors taken in the {\it
polynomial\/} form do rewrite, upon dressing with ghosts, as $\N4$
descendants. That they satisfy $\N4$ highest-weight conditions can be proven
in general, and therefore we conjecture a 1:1 correspondence between
$\SL2_{k=-4}$ singular vectors and those in the $(\N4)_{k=1}$ module. As
explained above, they give rise to cohomology in factor-modules.

\section{$\cQ^{(0)}$ cohomology by bosonization \label{CBB}}\lvm
We have observed the $\N4$ algebra in the cohomology of $\cQ^{(0)}$. It is
very useful to `straighten' the construction by explicitly projecting onto
the cohomology of $\cQ^{(0)}$.  This amounts to projecting out
(normal-ordered) operator monomials that contain $\cB^0$ and ${\hat J}^0$
(and their derivatives), since $\cB^0$ is not $\cQ^{(0)}$-closed while ${\hat
J}^0$ is $\cQ^{(0)}$-exact (see~\req{hatJ},\req{hatJexact}).  Effectively
solving the constraint ${\hat J}^0\sim0$ will now be achieved by using a
particular representation of the $s\ell(2)$ currents and will result in a
`strong' $N\!=\!4$ algebra (rather than the one modulo $\cQ^{(0)}$-exact
terms).

\subsection{Bosonizing $s\ell(2)$ currents and ghosts}\lvm
In this subsection, we will introduce a representation of the $\SL2$ currents
and the associated ghosts that would allow us to extract the $\N4$ algebra
at the level of operator products, not only in cohomology.

We start with the representation from ref.~\cite{[S-sing]} which is `induced'
from a Verma module of a minimal conformal theory.  Generally, this
representation can be considered as an embedding
$s\ell(2)\hookrightarrow\cM_d\oplus {\cal L}\oplus [B,C]\oplus U(1)_v$ where
$\cM_d$ denotes the minimal model, ${\cal L}$ is the theory of a free scalar
$\varphi$, $[B,C]$ is a ghost system, and $U(1)_v$ is an additional scalar
theory of a field  $v$ whose signature is opposite to that of $\varphi$.  The
minimal-model central charge $d$ is taken to satisfy the hamiltonian
reduction formula~\cite{[BO1]}
\BE
d=13 -6(k+2) - {6\over k+2}
\EE
where $k$ is the level of the $s\ell(2)$ algebra.  Setting, in our case,
$k=-4$, leads to $d=28$. Such a $d=28$-model is naturally viewed as a
`Liouville' theory; similarly, the scalar $\varphi$ that used to be a
`Liouville' in ref.~\cite{[S-sing]} is redefined by absorbing the imaginary
unit into it, which gives $\varphi$ a matter-like signature (and restores, in
our conventions, real background charge).  The signature of the additional
scalar $\d v$ changes similarly from matter-like in the formulae of
\cite{[S-sing]} to Liouville-like.  After all these redefinitions, the
formulae of ref.~\cite{[S-sing]} for $s\ell(2)_{-4}$ currents take the form
\BE\new\BA{rcl}
J^{+}&=&e^{\varphi-v},\qquad
J^0~{}={}~ BC - \d\varphi + 2\d v, \\
J^{-}&=&(T_m+\half\d\varphi\d\varphi-\threehalves\d^2\varphi-\d B C
-2B\d C + BC\d\varphi)e^{-(\varphi-v)}
\EA
\label{SemikhBos}
\EE
where
\BE
v(z)v(w)=-\log(z-w)\,,\qquad
\varphi(z)\varphi(w)=\log(z-w)\,.
\label{OPE2}
\EE
The ghost, $\varphi$- and $v$- \emt s read
\BE
T_{BC}=- B\d C\,,\qquad
T_\varphi=\half(\d\varphi)^2 - \threehalves\d^2\varphi\,,\qquad
T_v=-\half(\d v)^2 + \threehalves\d^2v
\EE
Then the currents $J^+$, $J^0$, and $J^-$ are given dimensions 0, 1, and 2
respectively with respect to the \emt\ $T_{BC}+T_\varphi+T_v+T_{\rm m}$.

In addition to the $s\ell(2)$ currents, the system of $T_{\rm m}$, $\varphi$,
$v$ and $BC$ ghosts allows one to represent an independent fermionic ghost
system, which we are going to identify with $\cB^-\cC_-$:
\BE
\cB^-=B\,e^{-(\varphi-v)},\qquad\cC_-=C\,e^{\varphi-v}
\label{BCbosonization}\EE
Evaluating the twisted Sugawara and $\cB^-\cC_-$ ghosts' \emt s in terms of
the `elementary' fields $T_{\rm m}$, $\varphi$, $v$ and $BC$, we arrive at
the relation
\BE
\widetilde{T^{\rm S}} - \cB^-\d\cC_-=
T_{\rm m} + \half(\d\varphi)^2 - \threehalves\d^2\varphi - B\d C
-\half(\d v)^2 + \threehalves\d^2v
\label{Tidentity}
\EE
which shows that we have an equal number of degrees of freedom in the
$\cM_d\oplus {\cal L}\oplus [B,C]\oplus U(1)_v$ system and in the
$s\ell(2)_{-4}$ currents with one ghost pair ($\cB^-\cC_-$).

In section~2, however, we had more ghost pairs.  Our next objective is to
extend the representation~\req{SemikhBos}, \req{BCbosonization} so as to
incorporate the other ghosts and then derive a realization for the $N\!=\!4$
algebra. This would require changing coordinates on the field space so as to
effectively solve the constraint ${\hat J}^0\sim0$.  To this end, we first
notice that since ${\hat J}^0$ is OPE-isotropic, ${\hat J}^0(z){\hat
J}^0(w)=0$, it can be represented by a complex scalar $\d\phi$, with
\BE
\phi(z)\bar\phi(w)=\log(z-w)\,,\quad
{\hat J}^0=\d\phi\,.
\EE
In the formulae \req{SemikhBos} and \req{BCbosonization}, we did have an
isotropic combination $\varphi-v$, and from the $s\ell(2)$ algebra we see
that this must be conjugate to $\d\phi={\hat J}^0$. We thus take $\bar\phi$
to be equal to $\varphi-v$, which allows us to have
$J^+=\exp\bar\phi$. \ Next we extend the field space by an independent ghost
system denoted by $bc$, and express $(\d v, \d\varphi, \cB^+,\cC_+)$ through
$(\d\phi, \d\bar\phi, b,c)$ via
\begin{eqnarray}
\d v&=&bc+\d\phi-\d\bar\phi,\\ \d\varphi &=&bc+\d\phi,\\
\cB^+&=&c\,e^{\bar\phi}\,,\qquad
\cC_+~{}={}~b\,e^{-\bar\phi}\label{BCplus}
\end{eqnarray}
after which the $s\ell(2)_{-4}$ currents take the form:
\begin{eqnarray}
J^{+}&=&e^{\bar{\phi}} \nonumber\\
J^{0}&=& bc + BC + \d\phi - 2\d \bar{\phi}
\label{BosSL2}\\
J^{-}&=&(T_{\rm m} + \half\d\phi\d\phi - \threehalves\d^2\phi + bc\d\phi
+ BC\d\phi - 2b\d c - \d b c - 2B\d C - \d B C + BCbc )
e^{-\bar{\phi}}\nonumber
\end{eqnarray}
while the $\cB^+\cC_+$ and $\cB^-\cC_-$ ghosts are now given by:
\BE\new\BA{rclcrcl}
\cB^-&=&Be^{-\bar{\phi}}\,,&{}&\cC_-&=&Ce^{\bar{\phi}}\,,\\
\cB^+&=&c\,e^{\bar\phi}\,,&{}&\cC_+&=&b\,e^{-\bar\phi}\,.
\EA\label{pmghosts}
\EE
Note that the $\cB^0\cC_0$ ghosts `decouple' -- they do not participate in
field redefinitions.

Using formulae \req{BosSL2}, \req{pmghosts}, we can now evaluate the
\emt~\req{EMT} in terms of the new fields $\phi,\bar\phi,bc,BC,T_{\rm m}$:
\BE
\cT_\cA=
\underbrace{T_{\rm m}-b\d c-2B\d C-\d B C}
+\d\phi\d\bar\phi -\d^2\phi - \cB^0\d\cC_0
\label{BOSEMT}
\EE
It follows that $\dim B=2$ while $\dim b=1$; \ $bc$ thus represent a $c=-2$
matter and together with $T_{\rm m}$ (which plays the r\^ole of a Liouville)
and the $BC$ ghosts make up a realization of the bosonic string.

\subsection{Constructing the states}\lvm
Let us see now how the representation space can be constructed for the
realization \req{BosSL2}, \req{pmghosts}.  The vacuum is found by translating
the vacuum determined by \req{sl2highest}, \req{sl2ghost} to the present
picture.  This gives for~\req{jdressed} (assuming $j>0$ for definiteness)
\BE
\ket{j}\tensor\underbrace{\ket{j+1}_+\tensor\ket{0}_0\tensor\ket{1}_-
}_{\cB\cC\ {\rm ghosts}}
=\underbrace{\ket{\Delta(r,s)}_{\rm m}\otimes\ket{-j-1}_{bc}\otimes\ket{1}_{BC}
}_{c=-2\ \rm bosonic\ string}
\otimes\ket{0}_{0}\tensor\ket{\{0,0\}}_{\phi\,\bar\phi}
\label{vacua}\EE
where $\ket{\{0,0\}}_{\phi\,\bar\phi}$ is the trivial vacuum in the complex
scalar theory.  The $bc$ and $BC$ ghost vacua from the RHS of \req{vacua} are
characterized by creation/annihilation conditions depending (for $bc$) on the
$s\ell(2)$ spin $j$.
\BE\new\BA{lclclcl}
c_n \ket{-j-1}_{b,c}&=&0\quad n\geq j+2\,,&{}&
b_n \ket{-j-1}_{b,c}&=&0\quad n\geq-j-1\,,\\
C_n \ket1_{B,C}&=&0\quad n\geq1\,,&{}&
B_n \ket1_{B,C}&=&0\quad n\geq0\,.
\label{caoper}
\EA\EE
while the conformal dimension of the primary matter state
$\ket{\Delta(r,s)}_{\rm m}$ is taken from the Ka\v{c} table for $d=28$:
\BE
\Delta(r,s)={\textstyle{1\over4}}(-\half(r^2-1)-2(s^2-1)-2rs+2)=
-\eighth\Bigl( (r+2s)^2-9\Bigr)\,,
\label{dimensions}\EE
which rewrites as
\BE
\Delta(r,s)=-\half j(r,s)(j(r,s)+3)
\label{dimmatter}
\EE
where $j(r,s)=j_+(r,s)$, i.e.\
\BE
j(r,s)={r-1\over2}+2{s-1\over2}\quad\hbox{$r$ and $s$ are integer}
\EE

For $j=j_-(r,s)$, we can do by analogy with the previous case, for example
the formula \req{vacua} rewrites as
\BE
\ket{j}\tensor\underbrace{\ket{0}_+\tensor\ket{0}_0\tensor\ket{-j}_-
}_{\cB\cC\ {\rm ghosts}}
=\underbrace{\ket{\Delta(r,s)}_{\rm m}\otimes\ket{0}_{bc}\otimes\ket{-j}_{BC}
}_{c=-2\ \rm bosonic\ string}
\otimes\ket{0}_{0}\tensor\ket{\{0,0\}}_{\phi\,\bar\phi}
\label{vacuamin}\EE
The formula \req{dimmatter} for the dimension of the matter state
$\ket{\Delta(r,s)}_{\rm m}$ remains the same.

\medskip

As mentioned after the formula~\req{dressfromjmin}, it is possible to build
the $\N4$ states $\ket{\half,\Delta}_{N=4}$ in terms of matter
$\ket{\Delta(r,s)}_{\rm m}$ and ghosts $BC$ and $bc$; the explicit formula
reads
\BE
\ket{\half,\Delta}_{N=4}=
\ket{\Delta(r,s)}_{\rm m}\otimes\ket{0}_{bc}\otimes\ket{1}_{BC}
\label{n4statesboson}
\EE
where $\Delta=\Delta(r,s)-1$, with $-1$ being accounted for by a $c$-ghost
contribution.

\medskip

The subspace built on
$\ket{\Delta(r,s)}_{\rm m}\otimes\ket{-j-1}_{bc}\otimes\ket{1}_{BC}$
can be thought of as the space of states of a non-critical bosonic string
with $c=-2$ matter.  We are going to consider it in more detail.

\subsection{Bosonization of spectral sequence and the LZ states}\lvm Now we
are going to translate the spectral sequence associated with the
decomposition \req{Qdecomposition} to the representation in terms of the
`elementary' fields $\phi,\bar\phi,bc,BC,T_{\rm m}$. These latter
have then to be
assigned the following degrees (cf.~\req{degrees}):
\BE\new\BA{l}
\deg T_{\rm m}=\deg\cC_0=\deg\cB^0=\deg\d\phi=\deg\d\bar\phi=0\,,\\
\deg C=-\deg B=\deg e^{-\bar\phi}=-\deg e^{\bar\phi}=1\,,\\
\deg b=-\deg c=2\,.
\EA\EE
The different parts \req{SPSEQU} of the BRST current now take the form
\BE\new\BA{l}
\cJ^{(0)}=\cC_0\d\phi\,,\\
\cJ^{(1)}=C(T_{\rm m}-b\d c)-CB\d C +\d^2 C-\d(Cbc)
+C(\half\d\phi\d\phi-\half\d^2\phi+\d\phi bc)-\d(\d\phi C)\,,\\
\cJ^{(2)}=b\,,\\
\cJ^{(3)}=bC\cB^0\,.
\EA\label{Bosdecomposition}\EE
We can thus project onto the cohomology of $\cQ^{(0)}$.  Observe that the
BRST current $\cJ^{(1)}$ and the \emt\ \req{BOSEMT} can be rewritten as
\BE\new\BA{l}
\cT_\cA=\cT_{\rm str} + [\cQ^{(0)},\,\cB^0\d\bar\phi-\d^2\cB^0]\,,\\
\cJ^{(1)}=\cJ_{\rm str} +
[\cQ^{(0)}\,,
\,\half\cB^0(C\d\phi+Cbc-2\d C)-\textstyle{3\over2}\d\cB^0]
\EA\EE
where fields in the cohomology can be taken as
\BE\new\BA{l}
\cT_{\rm str}=T_{\rm m}-b\d c-2B\d C-\d B C\,,\\
\cJ_{\rm str}=C(T_{\rm m}-b\d c)-CB\d C +\d^2 C-\d(Cbc)\,.\\
\EA\label{TQstring}\EE
These are identified with the \emt\ and the BRST current of a bosonic string
with a $c=-2$ matter represented by the $bc$ system.

Further, along with $\cT_\cA$ and $\cJ_\cA$, the $N\!=\!4$ generators
\req{GEN4ALGSL} can be projected onto the $\cQ^{(0)}$-cohomology.  This will
produce a `strong' $N\!=\!4$ algebra, i.e.\ relations~\req{SUPERALGSL} will
be satisfied exactly rather than modulo $\cQ^{(0)}$-exact terms as was the
case with the generators~\req{GEN4ALGSL}.  Namely, the $N\!=\!4$ generators
now take the form
\BE\new\BA{rclcrcl}
\cT&=&\cT_{\rm str}
\,,&{}&\cG^1&=&b\,, \\
J^+_{N=4}&=&Cb\,,&{}&\cG^2&=&B\,,\\
J^0_{N=4} &=&\half(bc-BC)\,,&{}&\bar\cG_1&=&
c(T_{\rm m}-B\d C)+bc\d c-\d{(cBC)}+\d^2 c\,,\\
J^-_{N=4} &=&Bc\,,&{}&\bar\cG_2&=&\cJ_{\rm str}
\EA\label{GEN4ALG}\EE
Here, $\cT=\cT_{\rm str}$ and  $\bar\cG_2=\cJ_{\rm str}$ are given by
eqs.~\req{TQstring}, while $\cG^2$ plays the r\^ole of a superpartner to
$\cT$.  The system described by the \emt\ $\cT_{\rm str}$ represents a $c=-2$
matter coupled to gravity, with $T_{\rm m}$ playing (in accordance with the
value of its central charge) the r\^ole of a Liouville.

Cohomology of the bosonized $\cQ^{(0)}$ operator coincides with the $\N4$
algebra representation built on the ($\N4$) highest-weight state obtained by
dressing the string vacuum $\ket{\Delta(r,s)}_{\rm
m}\otimes\ket{0}_{bc}\otimes\ket{0}_{BC}$ with the appropriate number of
ghosts, eq.~\req{n4statesboson}.

\medskip

The BRST-primitives of the states \req{generalmffn4} can now be written as
BRST-primitives w.r.t.\ the BRST operator $\cQ_{\rm str}= \oint\cJ_{\rm str}$.
An important fact is that the MFF singular vectors \req{generalmffn4} take in
our representation the form (see \cite{[S-sing]} where this claim was based on
using the (general form of) representation \req{BosSL2}, or \cite{[GP]} where
reductions of singular vectors were studied by other means)
\BE
({\rm singular\ vector\ of\ }T_{\rm m}{\rm \ minimal\ model})
\otimes({\rm ghost\ part})
\label{split}\EE
Their BRST-primitives then become, upon factorization over the module
generated by the null vector, the Lian-Zuckerman states in the theory of
$c=-2$ matter dressed with gravity. The ghost number of a LZ state obtained
in this way from an MFF vector $\ket{{\rm MFF}\{r,s\}}$ is equal to
$r-j-1=\half(r+1)-s$ for $r>j$ (where the ghost number of the $c=-2$ string
counts the number of $C$ operators minus the number of $B$ operators and is
zero for the $sl_2$-invariant ghost vacuum), and 0 for $r\leq j$. Such an
accumulation of the LZ states in the ghost number 0 is related to the
embedding pattern of the $c=28$ Virasoro Verma modules shown in \req{picture}
where, in the right column,  every module is embedded into all the higher
ones.

Moreover, the Verma module embedding diagrams then project as in
\req{picture}, where the right column shows Virasoro Verma modules and
dimensions of their ground states. Every Virasoro Verma module is embedded
into {\it all\/} the higher ones. For the $\SL2$ modules this is not so,
instead there exist two $\SL2$-modules (corresponding to two different spins
$j$) that project onto the same Virasoro module (modules on the same level in
\req{picture}).  Their embeddings add up to the `complete' set of embeddings
for the Virasoro Verma modules.

Our expression \req{generalmffn4} for the $\N4$ singular vectors now gives
rise, in view of~\req{split}, to an  expression for singular vectors in the
$c=28$ Virasoro Verma module. Rather curiously, writing these in a closed
form requires the introduction of ghost systems (which, as noted above,
decouple in the course of the evaluation of \req{generalmffn4} in terms of
the representation \req{GEN4ALG}).  The possibility to arrive at such a
representation for the Virasoro singular vectors rests on the fact that the
representation \req{BosSL2} for the $\SL2_{k=-4}$ currents (or, more
generally, the $\SL2_k$ representation from ref.~\cite{[S-sing]}), unlike
conventional `bosonizations', does not imply the vanishing of any singular
vector.

The explicit construction \req{generalmffn4} does not, however, give
all the singular vectors; as noted above, we have not continued
it to all possible values of $r$ and $s$
in a closed form. This leaves aside a
part of the $c=28$-singular vectors and the corresponding Lian-Zuckerman
states. For these we only have explicit constructions in lower-level
cases. In particular, the state $\ket{{\rm MFF}\{1,2\}}_\cA$ gives rise
to the ground ring generator $x$, and its explicit construction in terms of
the $N\!=\!4$ generators reads
\BE
x=BCb\,\Psi_{\rm m}-\half\d b\,\Psi_{\rm m} +\half b\,\d\Psi_{\rm m}
=((J^0_{N=4})_{-1}\cG^2_0 +\half\cG^2_{-1})\, (\bar\cG^1)_1\ket{1}_\cA\,,
\EE
where $\Psi_{\rm m}$ is the operator that corresponds to the state
$\ket{\Delta(1,2)}$ in the matter sector.  It would be interesting to obtain
this operator in a systematic way, by extending the
formula~\req{generalmffn4} along the lines of~\cite{[ST2]}.

\section{$N\!=\!4$ and matter{}${}+{}$gravity}\lvm
In this section we will comment on the relation of the $\N4$ algebra
represented in the cohomology of $\cQ^{(0)}$ to the known symmetries
of matter+gravity systems.
The $\N4$ algebra \req{GEN4ALG} contains two twisted $\N2$ subalgebras,
realized on a common \emt\ $\cT$ and $U(1)$ current $\cH=2J^0$:
\BE
\left\{\new\BA{rcl}
\cT_1&=&\cT\\
\cH_1&=&\cH\\
\cQ_1&=&\cG^1\\
\cG_1&=&\bar\cG_1
\EA\right.
\qquad{\rm and}\qquad
\left\{\new\BA{rcl}
\cT_2&=&\cT\\
\cH_2&=&\cH\\
\cQ_2&=&\bar\cG_2\\
\cG_2&=&\cG^2
\EA\right.
\label{subalgebras}
\EE
In the first of these algebras, we can bosonize one of the ghost pairs as
\BE
B=e^{i\varphi}\,,\qquad C=e^{-i\varphi}\label{IDENT}
\EE
and consider $\varphi$ as the Liouville scalar.
Then the construction
becomes the $k\!=\!-4$-case of the known $N\!=\!2$ representation \cite{[GS3]}
in terms of matter dressed with gravity:
\BE\new\BA{rcl}
\cQ_1&=&b\\
\cG_1&=&c(T-\half (\d\varphi)^2+{\ap+\am\over\sqrt{2}}\d^2\varphi)+bc\d c
+\kalphapl\d c\d\varphi+\half (1-2\alpha_+^2)\d^2c\\
\cH_1&=&-bc-\kalphapl\d\varphi\\
\cT_1&=&T-\half (\d\varphi)^2+{\ap+\am\over\sqrt{2}}\d^2\varphi-b\d c
\label{spin1}\EA\EE
where $\alpha_-=-\sqrt{k+2}$, $\alpha_+=-{1/\alpha_-}$ and, in our case, $k$
is set to $-4$. \ $T$ is the \emt\ of matter with central charge equal to
$1-{6(k+1)^2\over k+2}$ which becomes 28 when $k\!=\!-4$, in which case $T$
coincides with $T_{\rm m}$.

Similarly, the other twisted $N\!=\!2$ subalgebra from \req{subalgebras}
becomes, after the bosonization
\BE b=e^{i\varphi},\qquad c=e^{-i\varphi}\EE
the other $\N2$ realization from \cite{[GS3]}:
\BE\new\BA{rcl}
\cQ_2&=&C(T-\half (\d\varphi)^2+{\alpha_+-\alpha_-\over\sqrt 2}
\d^2\varphi)+BC\d C
+\kalphapl\d C\d\varphi+\half (1-2\alpha_+^2)\d^2C\\
\cG_2&=&B\\
\cH_2&=&BC+\kalphapl\d\varphi\\
\cT_2&=&T-\half (\d\varphi)^2+{\alpha_+-\alpha_-\over\sqrt 2}\d^2\varphi
-\d BC-2B\d C
\label{spin2}\EA\EE
evaluated at $k=-4$.

The two realizations~\req{spin1} and \req{spin2} are related by an involutive
automorphism of the twisted $N\!=\!2$ algebra \cite{[GS3]}.  Now we see that
this automorphism lifts to the automorphism \req{automorphism} of the $N=4$
algebra \req{GEN4ALG}.  It is induced, in the bosonized picture, by ghost
permutations (cf.~\req{pairs1})
\BE
b\leftrightarrow B\,,c\leftrightarrow C
\label{AUT}\EE
This acts as identity on the $s\ell(2)_{-4}$ algebra \req{BosSL2}.

More generally, one can construct a 3-parameter family of $N\!=\!2$ algebras
out of the $\N4$ generators \req{GEN4ALG} so that all members of this family
share the \emt\ $\cT$ and the $U(1)$ current $2J^0_{N=4}$.  These $N\!=\!2$
algebras are described by
\BE\new\BA{rcl}
\cT&=&T_{\rm m}-b\d c -2B\d C-\d BC, \\
\cH&=&bc-BC,\\
\cQ&=&a_1\bar\cG_2+a_2\cG^1,\\
\cG&=&a_3\cG^2+a_4\bar\cG_1
\label{family}\EA\EE
where $a_1,a_2,a_3,a_4$ are arbitrary parameters subject to
the equation
\BE
a_1a_3+a_2\,a_4=1\label{EQU}
\EE
Different algebras from the set \req{family} can be connected by a
combination of transformations of the form
$e^{-\oint\cA}(\ldots)e^{\oint\cA}$ with $\cA$ being equal to:

\BE\new\BA{rcl}
\cA&=&p_1bc-p_2BC\\
{\rm or}&&{}\\
\cA&=&p(T_{\rm m}C\,c-Cb\d c\,c-B\d C\,C\,c+C\d^2c)
\EA\EE
On the set of parameters $a_1,a_2,a_3,a_4$
these transformations act as
\BE\new\BA{rcl}
(a_1,a_2,a_3,a_4)&\to& (e^{p_2}a_1,e^{p_1}a_2,e^{-p_2}a_3,e^{-p_1}a_4)\\
{}{\rm and}\hfill &&{}\\
(a_1,a_2,a_3,a_4)&\to&(a_1+pa_2,a_2,a_3,a_4+pa_3)
\EA\EE
respectively.  As one can see,
not every two algebras of the family \req{family} can
be mapped into each other by such  a transformation.  The space of parameters
$a_1,a_2,a_3,a_4$ falls into three domains:
\BE\new\BA{rcl}
{}&(a_1,0,a_3,a_4)&{}\\
{}&(a_1,a_2,0,a_4)&{}\\
{}&(a_1,a_2,a_3,a_4)&{}
\EA
\label{threesets}
\EE
where in the latter case neither $a_2$ nor $a_3$ is equal to zero.  For an
algebra that belongs  to the first domain (that is $a_2=0$, hence $a_3\neq0$
from \req{EQU}) the cohomology of $\cG$ is trivial, because a
dimension-$(-1)$ field exists:
\BE
\Psi^{\cG}_1={1\over a_3}\,C-{a_4\over a_3^2}\,c\,\d C\,C
\EE
that is conjugate to $\cG$:
\BE
\cG(z)\Psi^{\cG}_1(w)={1\over z-w}
\EE
In the case of domain 2, we have a reversed situation.
A dimension-$0$ field:
\BE
\Psi^{\cQ}_2={1\over a_2}\,c-{a_1\over a_2^2}\,C\,\d c\,c\,,\qquad
a_2\neq0
\EE
is conjugate to $\cQ$, while the one conjugate to $\cG$ does not exist.  In
this case the cohomology of $\cQ$ are trivial.  In the last case (with
neither $a_2$ nor $a_3$ being zero) we have that $\Psi^{\cQ}_3=\Psi^{\cQ}_2$
as well as $\Psi^{\cG}_3=\Psi^{\cG}_1$ exist.  Hence the cohomology of both
$\cQ$ and $\cG$ are trivial.  Therefore the would-be transformations
connecting different domains do not exist, since such transformations are
required to preserve the OPEs.  The set of transformations of the form
$e^{-\oint\cA}(\ldots)e^{\oint\cA}$ act transitively on each domain.  The
above algebras \req{spin1} and \req{spin2} correspond, obviously, to
$(1,0,1,0)$ and $(0,1,0,1)$. At the same time, the $N\!=\!2$ algebra
\req{N2}, restricted to the cohomology of $\cQ^{(0)}$ in the bosonized
description, is identified as the $(1,1,\half,\half)$ algebra.

\medskip

The appearance of field operators such as the above $\Psi$ is
often characteristic to bosonized pictures; thus, for example,
the cohomology of $\cQ_\cA$ is trivialized in the bosonized
picture, since there exists a dimension-$0$ field \BE \Psi=c-C\d
c\,c+C\,c\,\cB^0 \label{Psi} \EE that is conjugate to the BRST
current $\cJ_\cA$:  \BE \cJ_\cA(z)\Psi(w)={1\over z-w} \EE
whence $\{\cQ_\cA,\,\Psi_0\}=1$ and we conclude that every $\cQ_\cA$ closed
state is $\cQ_\cA$ exact.  This vanishing of the cohomology is an example of
the Koszul trivialization, described e.g. in \cite{[A]}.  It occurs when
extending the algebra $\cA$ \req{A} to an algebra $\bar\cA$ and extending the
BRST operator appropriately, in such a way that any $\cQ_\cA$-closed state is
given by $\cQ_\cA$ acting on one of the `new' states.  The appearance of the
above $\Psi$ is due to explicitly solving the condition ${\hat J}^0\sim0$.
Indeed, taking ${\hat J}^0$ to be an `elementary' field and parametrizing the
fields orthogonal to ${\hat J}^0$ in terms of other `elementary' fields, one
has to introduce, one way or another, $(J^+)^{-1}$ (in our realization, this
was simply $e^{-\bar\phi}$). Allowing $(J^+)^{-1}$ to appear leads to the
existence of an operator such as the above $\Psi$ that trivializes the
cohomology.  To restore the `original' cohomology of $\cQ_\cA$, one has to
project from the bosonized algebra $\bar\cA$ to $\cA$ itself.  Recall that,
generally, the cohomology of $\cQ_\cA$ can be evaluated using the spectral
sequence associated with the decomposition \req{Qdecomposition}.  While the
cohomology of $\cQ^{(0)}$ and $\cQ^{(1)}=\cQ_{\rm str}$ do not necessarily
vanish, it is the cohomology of $\cQ^{(2)}=\oint\cG^1$ that undergoes the
Koszul trivialization due to the appearance of the $c$ field in the bosonized
picture. Note, however, that the $c$ ghost is {\it not\/} a part of the $\N4$
algebra, and thus it is possible to keep the non-trivial cohomology by
working solely in the representation of the $\N4$ algebra.

\section{Concluding remarks}\lvm
We have presented arguments showing that the $k=-4$ $\SL2$ WZW model is
cohomologically equivalent to the bosonic string with $c=-2$ matter.  We have
observed the presence of an $\N4$ symmetry in this system, which points to an
$\N4$ origin of the Lian--Zuckerman states in the $c=-2$ bosonic string.  The
derivation involves a spectral sequence on the $\SL2_{-4}$ BRST complex
and can be viewed as an extension of the Universal string ideology to include
theories with Ka\v{c}--Moody symmetries\footnote{With the help of the explicit
realization for the $\SL2$ currents (in terms of matter, Liouville and
ghosts) one can also construct a homotopy transformation relating the $\SL2$
space of states in the chosen realization with states of the matter+gravity
theories.}. While the $\N4$ algebra is specific to $c=-2$ matter, the
relation of Lian--Zuckerman states to the $\SL2_k$ algebra is likely to hold
in general, since a twisted $\N2$ is always present in non-critical strings,
and on the other hand the relevant $\N2$ singular vectors are isomorphic with
$\SL2$ singular vectors~\cite{[ST2]}.  The appearance of the $\N4$ algebra
would also be interesting to understand in terms of geometry of flag
manifolds, along the lines of ref.~\cite{[FS]}.

\bigskip

\noindent
{\sc Acknowledgements}.
We are grateful to B.~Feigin for useful discussions. We also thank
O.~Andreev, J.M.~Figueora-O'Farrill, S.~Hwang, O.~Khudaverdyan, A.~Marshakov,
A.~Taormina, I.V.~Tyu\-tin, M.A.~Vasil\-iev, and B.L.~Vo\-ro\-n\-ov. The
research described in this publication was made possible in part by Grant
\#MQM300 from the International Science Foundation and Government of Russian
Federation, and by RFFI grant 94-02-06338-a.


\begin{thebibliography}{99}
\addtolength{\baselineskip}{-.11\baselineskip}
\parindent=0pt
\parskip=-5pt

\def\NPB{Nucl. Phys. B}
\def\PLB{Phys. Lett. B}
\def\MPLA{Mod. Phys. Lett. A}
\def\CMP{Commun. Math. Phys.}
\def\IJMPA{Int. J. Mod. Phys. A}
\def\MPLA{Mod. Phys. Lett. A}
\def\JMP{J. Math. Phys.}

\bibitem{[AGSY]} O.~Aharony, O.~Ganor, J.~Sonnenschein, S.~Yankielowicz,
and N.~Sochen,
\NPB399 (1993) 527;\\
O.~Aharony, J.~Sonnenschein, and S.~Yankielowicz,
\PLB289 (1992) 309;\\
O.~Aharony, O.~Ganor, J.~Sonnenschein, and S.~Yankielowicz,
\NPB399 (1993) 560.
\bibitem{[HY]}H.-L.~Hu and M.~Yu,
\NPB391 (1993) 389.
\bibitem{[BMP]} P.~Bouwknegt, J.~McCarthy, and K.~Pilch {\sl Semi-infinite
cohomology in conformal field theory, and 2d gravity\/}, CERN-TH.6646/92.
\bibitem{[HR]}S.~Hwang and H.~Rhedin, {\sl Construction of BRST
invariant states in $G/H$ WZNW models\/}, G\"oteborg ITP 94-26 (December
1994).
\bibitem{[MFF]}F.G.~Malikov, B.L.~Feigin, and D.B.~Fuchs, Funk. Anal.
Prilozh. 20 no\,2 (1987) 25.
\bibitem{[LZ]}B.H.~Lian and G.J.~Zuckerman, \PLB254 (1991) 417.
\bibitem{[S-sing]}A.M.~Semikhatov,
\MPLA9 (1994) 1867.
\bibitem{[Distler]}J.~Distler \NPB342 (1990) 523.
\bibitem{[MV]}S.~Mukhi and C.~Vafa, \NPB407 (1993) 667;\\
S.~Mukhi, {\sl The Two-dimensional string as a topological field theory\/},
TIFR-TH-93-61.
\bibitem{[GS3]}B.~Gato-Rivera and A.M.~Semikhatov, Nucl. Phys. B408 (1993)
133.
\bibitem{[BLNW]}M.~Bershadsky, W.~Lerche, D.~Nemeschansky, and N.P.~Warner,
Nucl. Phys. B401 (1993) 304.
\bibitem{[BV]}N.~Berkovits and C.~Vafa, \MPLA9 (1994) 653.
\bibitem{[Fof]}J.M.~Figueora-O'Farrill, \PLB321 (1994) 344.
\bibitem{[IK]}H.~Ishikawa and M.~Kato, \MPLA9 (1994) 725.
\bibitem{[KSch]}D.~Karabali and H.~Schnitzer, \NPB329 (1990) 649.
\bibitem{[ISRA]}J.M.~Isidro and A.V.~Ramallo \PLB316 (1993) 488.
\bibitem{[Frenkel]}B.~Feigin and E.~Frenkel, \IJMPA7, Suppl. 1A (1992) 197.
\bibitem{[FMS]}D.H.~Friedan, E.J.~Martinec, and S.H.~Shenker, \NPB271 (1986)
93.
\bibitem{[Kazama]}Y.~Kazama, \MPLA6, 14, 1321 (1991).
\bibitem{[Ey]}T.~Eguchi and S.-K.~Yang, \MPLA4 (1990) 1653.
\bibitem{[W-top]}E.~Witten, Commun. Math. Phys. 118 (1988) 411; \NPB 340
(1990) 281.
\bibitem{[Mats]} S.~Matsuda, KUCP-42 (December 1991).
\bibitem{[ET]}T.~Eguchi and A.~Taormina, \PLB196 (1986) 75.
\bibitem{[PT]}J.L.~Petersen and A.~Taormina,  {\sl Coset construction and
character sumrules for the doubly extended $N=4$ superconformal algebras\/},
DTP/92/49, NBI-HE-92-73 (November  1992).
\bibitem{[BO1]}M.~Bershadsky and H.~Ooguri,
Commun. Math. Phys. 126 (1989) 49.
\bibitem{[GP]}A.Ch.~Ganchev and V.B.~Petkova, Phys. Lett. B318 (1993) 77;
\PLB293 (1992) 56; P.~Furlan, A.Ch.~Ganchev, R.~Paunov and V.B.~Petkova,
Nucl. Phys. B394 (1993) 665.

\bibitem{[BS]}M.~Bauer and N.~Sochen, {\sl Fusion and singular vectors in
$A_1^{(1)}$ highest weight cyclic modules\/}, {\tt hep-th@xxx}/9201079.

\bibitem{[FS]}B.L.~Feigin and A.V.~Stoyanovsky, {\sl
Quasi-particle models for the representations of Lie algebras and
geometry of flag manifold\/}, RIMS-942 (Sept. 1993).

\bibitem{[SchS]}A.~Schwimmer and N.~Seiberg, \PLB184 (1987) 191.

\bibitem{[ST2]}A.M.~Semikhatov and I.Yu.~Tipunin, {\sl Singular Vectors of
the Topological Conformal Algebra\/}, {\tt hep-th@xxx/9512079}.

\bibitem{[A]}F.~Akman, {\sl Some cohomology operators an 2-D field theory\/},
MSRI preprint, and PhD thesis, Yale.

\end{thebibliography}
\end{document}